\documentclass[1p, 12pt]{elsarticle}

\usepackage[export]{adjustbox}
\usepackage[utf8x]{inputenc}
\usepackage[english]{babel}
\usepackage[T1]{fontenc}
\usepackage{lmodern}
\usepackage{epstopdf}
\expandafter\let\csname equation*\endcsname\relax
\expandafter\let\csname endequation*\endcsname\relax
\usepackage{amsmath,amsfonts,amssymb,amsthm,bm,times,dcolumn}
\usepackage[normalem]{ulem}
\usepackage{microtype}
\usepackage{braket}
\usepackage{gensymb} % for example \degree symbol can be used!
\usepackage{physics}
\usepackage[title]{appendix}

\usepackage[colorlinks={true}, citecolor={blue}, filecolor={blue}, linkcolor={blue}, urlcolor={blue}]{hyperref}
\usepackage{graphicx,color}
\usepackage[caption=false]{subfig}
\begin{document}

%\preprint{APS/123-QED}

\title{Entanglement transfer via Chiral and Continuous-Time Quantum Walks on a Triangular Chain}

\author{Utku Sa\u{g}lam}
\ead{usaglam@wisc.edu}
\address{Department of Physics, 1150 University Avenue, University of Wisconsin at Madison, Madison, Wisconsin 53706, USA}

\author{Mauro Paternostro}
\ead{m.paternostro@qub.ac.uk}
\address{Centre for Theoretical Atomic, Molecular, and Optical Physics, School of Mathematics and Physics,
Queen’s University, Belfast BT7 1NN, United Kingdom}

 \author{\"{O}zg\"{u}r E.\ M\"{u}stecapl{\i}o\u{g}lu}
\ead{omustecap@ku.edu.tr}
\address{Department of Physics, Ko\c{c} University, Sar{\i}yer, \.Istanbul, 34450, Turkey}
\address{
T\"{U}BİTAK Research Institute for Fundamental Sciences, 41470 Gebze, Turkey}

\date{\today}

\begin{abstract}
We investigate the chiral quantum walk (CQW) as a mechanism for an entanglement transfer on a triangular chain structure. We specifically consider two-site spatially entangled cases in short-time and long-time regimes. Using the concurrence as an entanglement measure; fidelity, and the Bures distance as the measure of the quality of the state transfer, we evaluate the success of the entanglement transfer. We compare the entangled state transfer time and quality in CQW against a continuous-time quantum random walk. We also observe the effect of mixed states on the entanglement transfer quality.
\end{abstract}

\maketitle

%\tableofcontents
%%%%%%%%%%%%%%%%%%%%%%%%%%%%%%%%%%%%%%%%%%%%%%%%%%%%%%%%%%%%%%%%%%%%%
\section{\label{sec:intro}Introduction}
%%%%%%%%%%%%%%%%%%%%%%%%%%%%%%%%%%%%%%%%%%%%%%%%%%%%%%%%%%%%%%%%%%%%%

Fast and accurate transmission of quantum states through communication or computation networks is a critical objective for quantum technologies~\cite{bennett_quantum_2000,divincenzo_physical_2000,kimble_quantum_2008}. Proposed schemes to achieve this goal consider engineered couplings between the network sites~\cite{bose_quantum_2003,christandl_perfect_2004,albanese_mirror_2004,wojcik_unmodulated_2005,di_franco_perfect_2008,mohseni_environment-assisted_2008,apollaro_99-fidelity_2012,sinayskiy_decoherence-assisted_2012,korzekwa_quantum-state_2014,estarellas_robust_2017}, external fields~\cite{shi_quantum-state_2005,hartmann_excitation_2006,banchi_optimal_2010,shan_controlled_2018}, weak measurements~\cite{he_robust_2013,man_controllable_2014}, or transport in noisy environments of biological or synthetic systems~\cite{zwick_optimized_2014,li_one-way_2019,matsuzaki_one-way_2020,vieira_almost_2020,PhysRevResearch.2.013369}. Such methods are challenging to implement in practice for quantum entanglement transfer, due to quantum decoherence and disorder~\cite{de_chiara_perfect_2005,ashhab_quantum_2015}.

Pretty-good State Transfer (PGST) can be achieved on dual-spin chains ~\cite{DualRail}, spin chains with weakly coupled endpoints \cite{wojcik_unmodulated_2005,PhysRevA.76.052328,Banchi_2011,PhysRevA.78.022325,Giampaolo_2010}, and projective measurements ~\cite{measurementAssist} with Quantum Walk (QW) schemes. Continuous time quantum walk (CTQW) is a paradigmatic model of quantum transport~\cite{MULKEN201137,alg}. Both discrete-~\cite{zhan_perfect_2014,stefanak_perfect_2016} and continuous-time~\cite{kendon_perfect_2011,large_perfect_2015,Cameron2014,Connelly2017} quantum walks have been discussed for PGST. 
CTQW can be made one way by taking complex-valued couplings, which is called a chiral quantum walk (CQW)~\cite{Zimbors2013,PhysRevA.93.042302,Wong2015}. Chirality emerges due to the breaking of time-reversal symmetry (TRS)~\cite{liu_quantum_2015}, and it provides a significant boost to transport speed~\cite{Zimbors2013}.

High-dimensional entanglement (entanglement in high-dimensional degrees of freedom, such as spatial path modes) is advantageous in quantum communication~\cite{Cozzolino2019} and quantum superdense coding~\cite{6Liu2002,7Grudka2002,8Hu2018}. The perfect state transfer (PST) in spin chains paves the way for the creation of required entangled states and logic gate structures for quantum computation and quantum information ~\cite{kay_review, Kay2005}. High-dimensional entangled states can be produced by repeatedly generating the entanglement in a low-dimensional system and transferring these to a higher dimensional one~\cite{Giordani2021}. Our first goal is to explore if CQW can be used to transmit two-dimensional quantum entangled states; what are the possible advantages it may offer? In addition,
we ask if and to which extent CTQW can be used in place of CQW with the same chiral properties. CTQW can be easier to implement than CQW. For that aim, we identify the underlying physics of the chiral nature of QW in terms of quantum path interference,
which can be controlled either via the phase in the initial state for CTQW or via the 
phase in complex hopping coefficients in CQW.

We specifically consider CQW on a linear spin chain of equilateral triangles, as shown in Fig.~\ref{fig:triChain}, which is the simplest graph that allows for so-called probability time symmetry (PTS) breaking~\cite{PhysRevA.93.042302}. A walker can transfer from one node to any other neighboring site on the triangular plaquette by passing through either one or two edges. We consider a uniform complex coupling between the nearest-neighbor sites.
Due to the path length difference between the odd and even number of edges traveled, and phases of the complex couplings, interference can enhance the transfer rate. Path interference in the context of quantum walk means that the relative phase between different trajectories the particle can
traverse from one site to another one can give constructive or destructive interference effects in the site-to-site probability transfer. By using special graph topologies, complex-valued site-to-site couplings, or specific initial states, one can use path interference to break the PTS. CTQW can only utilize the latter, initial states with
specific phases to exploit quantum interference while CQW can use both the freedom to choose the initial state and complex hopping coefficients to break PTS.

The spatial entanglement can be defined in the site basis for quantum walks \cite{spatialEntanglementQW, singleParticleEnt}. We assume a particle (we call a spin excitation as a particle)  in a Bell-like spatially entangled state of two sites injected into the chain from the left. The quality of the transfer is examined by calculating the density matrix, concurrence~\cite{PhysRevLett.80.2245}, fidelity, and Bures distances explicitly~\cite{Jozsa1994,Hbner1992,Hbner1993}. We have also numerically confirmed that the entanglement state transfer time linearly scales with the chain size \cite{tony1, tony2, tony3}.

The triangular chain lattice can be realized in superconducting circuits~\cite{Vepslinen2020,Ma2020}, trapped ions~\cite{trappedIonChain}, NMR systems~\cite{PhysRevA.93.042302}, photonic and spin waveguides~\cite{PhysRevA.93.062104}, and in optical lattices~\cite{PhysRevLett.93.056402}. In the case of optical lattices, complex edge weights could be introduced with the help of artificial gauge fields~\cite{bloch_many-body_2008,aidelsburger_artificial_2018}, nitrogen-vacancy centers in diamonds ~\cite{levi} or with plasmonic non-Hermitian coupled 
waveguides~\cite{Fu2020}.

This paper is organized as follows. We introduce the CQW on a triangular chain model by presenting the adjacency matrix and present the associated Hamiltonian model with complex hopping rates in Sec.~\ref{Sec:Model}.
Our results are given in Sec.~\ref{Sec:Results} in five subsections.
PTS breaking and entanglement transfer in CTQW and CQW on a triangular chain are discussed in Sec.~\ref{Sec:PTSB-CTQW} and Sec.~\ref{Sec:PTSB-CQW}, respectively. We conclude in Sec.~\ref{Sec:Conclusion}.

%%%%%%%%%%%%%%%%%%%%%%%%%%%%%%%%%%%%%%%%%%%%%%%%%%%%%%%%%%%%%%%%%%%%%
\section{CQW on a triangular chain}\label{Sec:Model}
%%%%%%%%%%%%%%%%%%%%%%%%%%%%%%%%%%%%%%%%%%%%%%%%%%%%%%%%%%%%%%%%%%%%%

%%%%%%%%%%%%%%%%%%%%%%%%%% FIGURE 1 %%%%%%%%%%%%%%%%%%%%%%%%%%%%%%%%%
\begin{figure}[t!]
	\label{fig1}
	\centering
	\includegraphics[width=\linewidth]{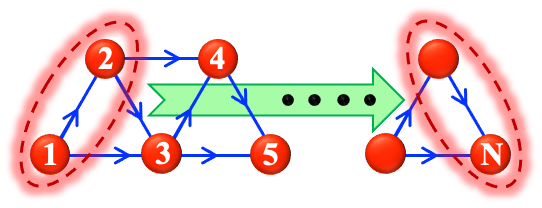}
	\caption{\label{fig:triChain} Graph of a linear chain with $N$ vertices arranged in triangular plaquettes. Initially, the pair of sites $1$ and $2$ at the left end of the chain are entangled.
	The entangled state is transported to the rightmost pair of vertices ($N-1$ and $N$) after a chiral quantum walk. Arrows indicate the directed edges with complex weight factors, taken to be $+i$. In the opposite direction, the weight factors change phase and become $-i$. We have also examined transferring three-site entanglement from the leftmost plaquette ($1,2$ and $3$) to the right end of the chain.}
\end{figure}
%%%%%%%%%%%%%%%%%%%%%%%%%%%%%%%%%%%%%%%%%%%%%%%%%%%%%%%%%%%%%%%%%%%%%

Typical quantum walks exhibit time-reversal symmetry (TRS) in  transfer probabilities between sites $n$ and $m$ in forward ($t$) and backward ($-t$) times such that $P_{nm}(t) = P_{nm}(-t)$. CQWs break the TRS and allow for so-called “directionally biased” transport, $P_{nm}(t) \neq P_{mn}(t)$, in certain graph structures~\cite{PhysRevA.93.042302}.
We consider CQW on a triangular chain of $N$ vertices as shown in Fig.~\ref{fig:triChain}, which is a minimal configuration with PTS breaking for a quantum walk with a directional bias~\cite{PhysRevA.93.042302}.

We will use the site basis $\{|i\rangle\}$, with $i=1,...,\text{N}$ indicating which site is occupied such that $|i\rangle:=|0_1,0_2,..,1_i,..0_\text{N}\rangle$. The set of coupled sites in a graph determines the edges $e=(i,j)$, which can be described by the so-called adjacency matrix $A$~\cite{Kempe2003,Sett2019,
PhysRevA.58.915,Childs2002}. For a triangular chain of $N=5$ sites, $A$ is given by
\begin{equation}\label{eq:A}
A=
\begin{bmatrix}
0&1&1&0&0\\
1&0&1&1&0\\
1&1&0&1&1\\
0&1&1&0&1\\
0&0&1&1&0\\
\end{bmatrix}.
\end{equation}
Together with the degree matrix (For definitions of some graph theory terms see Appendix \ref{sec:footNotes}) $D$ for the self edges (i,i), $A$, determines the graph Laplacian $L=D-A$. Hamiltonian of the walk is given by the Hadamard product mentioned in the Appendix \ref{sec:footNotes}, $H=J\circ L$, where $J$ is the matrix of hopping rates (edge weights). We neglect the self energies; therefore, we will take $D=0$ and
write the Hamiltonian as
\begin{equation}\label{eq:H}
H=\sum_{nm}(J_{nm}A_{nm}|n\rangle\langle m|+J_{mn} A_{mn}|m\rangle\langle n|).
\end{equation}
In contrast to CTQW where every $J_{nm}$ is real-valued, CQW allows for complex edge weights, subject to $J_{mn}=J_{nm}^\ast$, so that the support graph of the walk becomes a directed one (cf.~Fig.~\ref{fig:triChain}). 
Specifically, we take
\begin{equation} 
\label{eq:Hcqw}
H=
\begin{bmatrix}
0&-\text{i}&-\text{i}&0&0\\
\text{i}&0&-\text{i}&-\text{i}&0\\
\text{i}&\text{i}&0&-\text{i}&-\text{i}\\
0&\text{i}&\text{i}&0&-\text{i}\\
0&0&\text{i}&\text{i}&0\\
\end{bmatrix}
\end{equation}

The choice of the phase $\theta_{nm} =\theta=\pi / 2$ of complex hopping weights $J_{nm}=|J_{nm}|\exp(\text{i}\theta_{nm})$ (with $n>m$) is based upon the general investigations of CQW~\cite{Zimbors2013} for a triangular plaquette. It is found that maximum bias in time asymmetry can be obtained at $\theta =\pi / 2$~\cite{Zimbors2013}. Remarkably, the spectrum of Hamiltonian with a phase of $\pi/2$ has an anti-symmetric structure
\begin{equation}\label{eq:specT}
\begin{split}
\Lambda_{1,2,3,4,5}&=
(-\sqrt{\frac{1}{2} (7 + \sqrt{37})}, -\sqrt{\frac{1}{2}(7 - \sqrt{37})},0,\\
&\sqrt{
\frac{1}{2} (7 - \sqrt{37})}, \sqrt{\frac{1}{2}(7 + \sqrt{37})}).
\end{split}
\end{equation}
We intuitively assume that a similar choice should yield efficient entanglement transfer along a linear chain of equilateral triangles, too. We numerically examined different choices and verified that our intuition is correct (Some typical results will be given in Sec.~\ref{Sec:Results}). The eigenstates corresponding to $\theta=\pi / 2$ are given in Appendix~\ref{sec:eigenstates}.

The evolution of the initial state of the system $\rho (0)$ is given by $\rho(t)=U \rho(0)U^\dag$ where $U:= \exp(\text{-i}Ht)$. We define the site states of the chain as $\ket{1},...\ket{i}$, where $i$'s are the site numbers. Therefore, as the initial state, we  consider a spatially entangled state $\ket{\psi_\text{spatial}}=(\ket{1}-\text{exp}(i\phi)\ket{2})/\sqrt{2}$ with the phase $\phi$, of the leftmost sites of the chain and we aim to transfer the state to the right end of the chain.

\begin{figure}[t!]
	\centering
	\includegraphics[width=\linewidth]{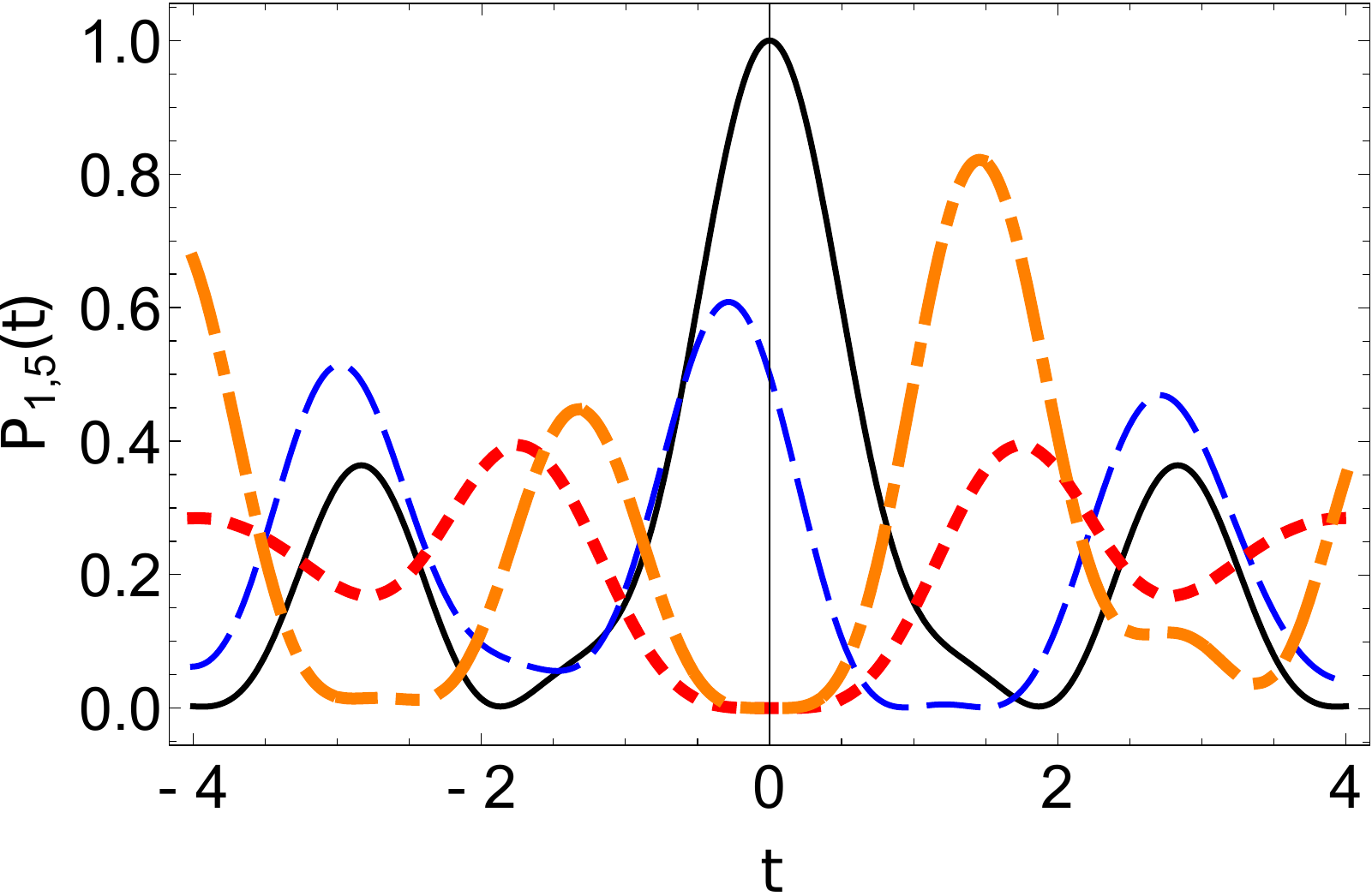}
	\caption{\label{fig:TRS-CTQW} Time dependence of the occupation probabilities  \(P_{1,5}(t)\) of sites $1$ and $5$ of a triangular chain with $5$ sites, where a particle makes CTQW. For an initially localized particle with an initial state \(\ket{1}\), $P_1(t)$ and $P_5(t)$ are indicated by black, solid, and dotted, red curves, respectively. For a particle initially in a superposition state $\ket{\psi(0)}= (\ket{1}-\exp(\text{i}\phi)\ket{2})/\sqrt{2}$ with  \(\phi=3\pi \slash 4\), $P_1(t)$ and $P_5(t)$ are shown as dashed, blue and dot-dashed, orange curves, respectively.}
\end{figure}

Remarkably, even with initial superposition states on our linear triangular chain, could not yield PST (cf.~Fig.~\ref{fig:TRS-CTQW}).
The graphs that can support PST require to be hermitian, circulant (\ref{sec:footNotes}), and to have a non-degenerate spectrum, together with a flat eigenbasis~\cite{Cameron2014}. Definitions of the graph theory terms we use are given in Appendix~\ref{sec:footNotes}. Alternatively, PST can still be achieved with a non-circulant graph that contains non-zero values on certain off-diagonal elements of
its adjacency matrix~\cite{Connelly2017}. From a practical point of view, implementing graph structures that have PST is challenging because of the sophisticated and usually numerous special connectivities of these graphs. Therefore, creating a simpler graph with PGST can be more feasible in practice than using a graph with a PST.
Since our proposed adjacency matrix is neither circulant nor has required non-zero off-diagonal elements, we do not expect any PST.

A fundamental difference between CQW and CTQW regarding the directional bias lies in how the transport bias is introduced. In CQW, the directional bias emerges by the differences in transition probabilities depending on the Hamiltonian regardless of the initial states to be transported. In CTQW, directional symmetry breaking is sensitive to the phase differences in the initial particle state in the (spatial) site basis.  Intuitively, there is an interplay of path interference and the initial phase in CTQW in breaking PTS. The significance of the difference between the CQW and CTQW in such a directionally biased entanglement transfer is the ability of CQW to break PTS for any initial condition, while directionally biased entanglement transfer in CTQW happens only for certain initial states.

In this paper, we consider two time ranges to investigate the entanglement transfer dynamics. The first one, which we call the short-time regime, is to probe the first maximum of the entanglement measure (concurrence) or success fidelity of the state transfer. The second case is called the long-time regime, allowing multiple scatterings of the particle at the ends of the chain. The latter case is used to probe if more successful entanglement transfer is possible or not, at the cost of longer transfer times.

%%%%%%%%%%%%%%%%%%%%%%%%%%%%%%%%%%%%%%%%%%%%%%%%%%%%%%%%%%%%%%%%%%%%%
\section{Results and Discussion}
\label{Sec:Results}
%%%%%%%%%%%%%%%%%%%%%%%%%%%%%%%%%%%%%%%%%%%%%%%%%%%%%%%%%%%%%%%%%%%%%

%%%%%%%%%%%%%%%%%%%%%%%%%%%%% FIGURE 2 %%%%%%%%%%%%%%%%%%%%%%%%%%%%%%
%%%%%%%%%%%%%%%%%%%%%%%%%%%%%%%%%%%%%%%%%%%%%%%%%%%%%%%%%%%%%%%%%%%%%

We will start with an examination of PTS-breaking in CTQW on the triangular chain.
We do not need complex edge weights to break PTS in general. The essential physical mechanism behind PTS breaking is the quantum path 
interference, for which
the required phase difference, or quantum coherence, can be injected into the initial state instead of the edges. 
Specifically, we consider a particle that is spatially entangled in two sites in the site basis,
\begin{eqnarray}\label{eq:ctqw-phaseInitialState}
\ket{\psi_{\text{spatial}}(0)} = \frac{1}{\sqrt{2}}(\ket{1}-\text{e}^{\text{i}\phi}\ket{2})
\end{eqnarray}
on a regular triangular chain with real-valued hopping weights (we take $J_{nm}=1$ for simplicity).

We want to transfer the input state to the end-sites of the structure. Ideally, we would like to achieve the target state $\ket{\psi_{\text{target}}}=(\ket{4}-\text{exp}(\text{i}\phi)\ket{5})/\sqrt{2}$. To characterize the performance of the actual process,  
in addition to the fidelity of state transfer $|\bra{\psi_{\text{target}}}\ket{\psi (t)}|^2$, we quantify the entanglement transferred to the end of the chain (sites $4$ and $5$). As we have only a single excitation, the initial state in Eq.~(\ref{eq:ctqw-phaseInitialState}) evolves into a state in the form
\begin{equation}
\begin{split}
\ket{\psi(t)}&=\left(A_1\ket{100}_{123}+A_2\ket{010}_{123}+A_3\ket{001}_{123}\right)\ket{00}_{45}\\
&+\ket{000}_{123}\left(A_4\ket{10}_{45}+A_5\ket{01}_{45}\right),\\
\end{split}
\end{equation}
where $A_i$ are the time-dependent coefficients depending on the eigenvalues of the Hamiltonian of the QW. 

Tracing out the states of the sites $1,2$ and $3$ in the density matrix $\rho(t)=\ket{\psi(t)}\bra{\psi(t)}$, we find the reduced density matrix in the computational basis $\ket{4}\ket{5}=\ket{00}$, $\ket{01}$, $\ket{10}$, $\ket{11}$ for the sites $4$ and $5$ in the form 
\begin{equation}
\label{eq:anDenMat}
\rho_{4,5}=
\begin{bmatrix}
1-a_{44}-a_{55}&0&0&0\\
0&a_{44}&a_{45}&0\\
0&a^*_{45}&a_{55}&0\\
0&0&0&0\\
\end{bmatrix},
\end{equation}
where $a_{ij}=A_i A^*_j$. The distribution of the zero elements, and hence the sparsity of the matrix, remains the same for any chain length.

We can quantify the pairwise entanglement using  
concurrence ~\cite{PhysRevLett.80.2245}. For a state $\rho$, concurrence is defined as
\begin{equation} \label{eq:conc}
C(\rho)=\text{max}(0,\lambda_{1}-\lambda_{2}-\lambda_{3}-\lambda_{4}),
\end{equation}
where $\tilde{\rho}$ is the spin-flipped state, and $\{\lambda_i\}$ is the set of eigenvalues of $R=(\rho^{1/2} \tilde{\rho}\rho^{1/2})^{1/2}$ arranged in non-increasing order. With this at hand, $C_{i=4,j=5}$ is found to take the form
\begin{equation}
\label{eq:concAnalytic}
C_{4,5}=2\text{Max}\left(0,\sqrt{a_{44}a_{55}},|a_{45}|\right).
\end{equation}
Owing to the definition of $a_{ij}$, the concurrence $C_{4,5}$ can further be simplified to $C_{4,5}=2|a_{45}|$. Note that this result can be extended to the matrices with arbitrary size associated with chains of length $N$, giving $C_{N-1,N}=2|a_{N-1,N}|$.

Using the Hamiltonian matrix $H$ in Eq.~(\ref{eq:Hcqw}) and the initial state  $\rho(0)=\ket{\psi_{\text{spatial}}(0)}\bra{\psi_{\text{spatial}}(0)}$ in  Eq.~(\ref{eq:ctqw-phaseInitialState}), with any $\phi$, the dynamics of the entanglement between pairs of sites of the system can be calculated numerically for any value of $t$.  We will also investigate the transport of the initial entangled Bell site-state from the sites $1$ and $2$ to sites $4$ and $5$ in CQW.
We will perform the calculations for both CQW and CTQW cases separately. In the case
of CTQW we use the initial state in Eq.~(\ref{eq:ctqw-phaseInitialState}).

%%%%%%%%%%%%%%%%%%%%%%%%%%%%%%%%%%%%%%%%%%%%%%%%%%%%%%%%%%%%%%%%%%%%%
\subsection{PTS breaking and entanglement transfer in CTQW on a triangular chain}
\label{Sec:PTSB-CTQW}
%%%%%%%%%%%%%%%%%%%%%%%%%%%%%%%%%%%%%%%%%%%%%%%%%%%%%%%%%%%%%%%%%%%%%

To appreciate the role of the initial phase in Eq.~(\ref{eq:ctqw-phaseInitialState}) on the state transfer and 
PTS breaking in CTQW on the triangular chain, let's start with the initial state \(\ket{\psi(0)}=\ket{1}\). The occupation probabilities 
$P_{i}= \langle i | \rho(t) | i\rangle$ of the sites \(i=1\) (solid, black) and \(i=5\) (dotted, red) are shown in Fig.~\ref{fig:TRS-CTQW}, where mirror symmetry in the behaviour of the probability distribution with respect to time can be seen. Transfer from the initially occupied site $\ket{1}$ to the rightmost site $\ket{5}$ is found to be weak (less than $45\%$ at any time). 

If we use the initial state given in Eq.~(\ref{eq:ctqw-phaseInitialState}) with the Adjacency matrix  Eq.~(\ref{eq:A}), in addition to being able to control path interference, PTS can be broken depending on the initial phase $\phi$. We have numerically compared $P_5(t)$ for different $\phi$
and found that $\phi=3\pi/4$ gives the largest occupation of site $|5\rangle$. Fig.~\ref{fig:TRS-CTQW} shows $P_1(t)$ (dotted, red) and $P_5(t)$ (dot-dashed, orange) for $\phi=3\pi/4$, where time reversal asymmetry, $P(t)\neq P(-t)$ emerges. 
Population transfer is
significantly enhanced using such a superposition state initially.  
We conclude that transferring a particle from the left end of the chain to a site at the right end is more successful by injecting the particle simultaneously at two sites with a certain quantum
coherence relative to starting a well-localized particle at a single site. Let's now explore if similar advantages can be found in the entanglement transfer.

%%%%%%%%%%%%%%%%%%%%%%%% FIGURE 3 %%%%%%%%%%%%%%%%%%%%%%%%%%%%%%%%%%%%%%%%%%%%%%%%%%
\begin{figure}[!t]
	\centering
	{\includegraphics[width=\linewidth]{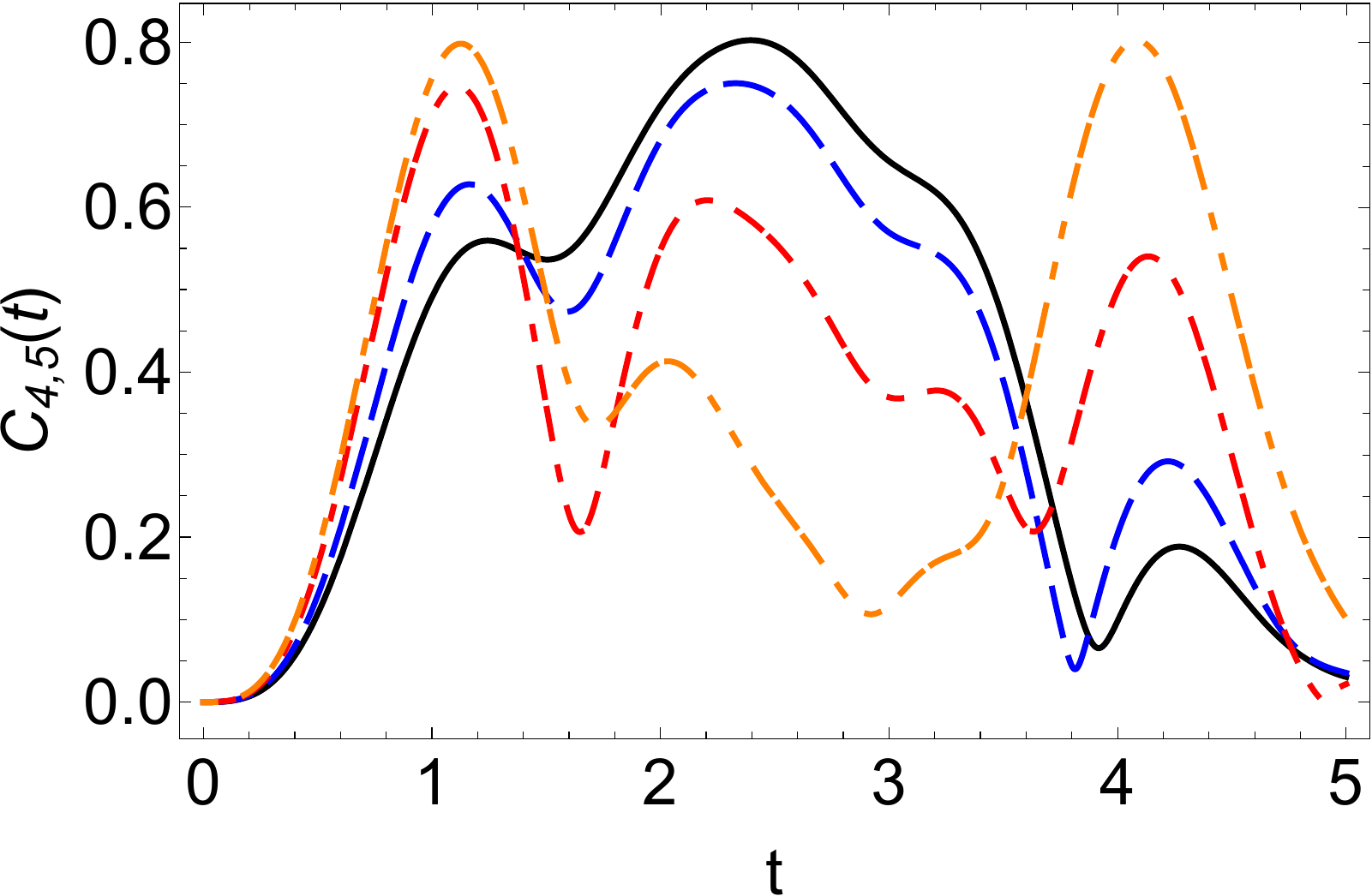}}
	\caption{\label{fig:CONC-CTQW}  Time dependence of the concurrence \(C_{4,5}(t)\) in the state of sites \(\ket{4}\) and \(\ket{5}\) for the initial state in Eq.~(\ref{eq:ctqw-phaseInitialState}) of a particle that makes CTQW on a triangular chain of $N=5$ sites. Different curves stand for the initial states
	$\ket{\psi(0)}= (\ket{1}-\exp(\text{i}\phi)\ket{2})/\sqrt{2}$ with
	\(\phi = \pi \slash 4\) (solid, black),  \(\phi = \pm \pi \slash 3\) (dashed, blue), \(\phi = \pm \pi \slash 2\) (dot-dashed, red), \(\phi = \pm 3 \pi \slash 4\) (dotted, orange). }
\end{figure}
%%%%%%%%%%%%%%%%%%%%%%%%%%%%%%%%%%%%%%%%%%%%%%%%%%%%%%%%%%%%%%%%%%%%%%%%%%%%%%%%%%%

Fig.~\ref{fig:CONC-CTQW} shows that the concurrence is optimum for \(\phi=\pm3\pi\slash4\) with a value \(C_{4,5}(1.12)\sim0.8\). Therefore, for pairwise entanglement transfer, \(\phi=3 \pi \slash 4\) gives the most advantageous initial state. A natural question to ask is if there is a fundamental connection between the critical phase \(\phi=3 \pi \slash 4\) and PTS breaking in CTQW.
 
We can quantify the bias between the forward and backward time evolutions using the
Bures distance between the states \(\rho(t)\) and \(\rho(-t)\). Bures distance is 
defined by~\cite{Jozsa1994,Hbner1992,Hbner1993}
\begin{equation}\label{bures}
D_{B}(\rho,\sigma)^{2}=2(1-\sqrt{F(\rho,\sigma)}),
\end{equation}
where
\begin{equation}
\label{eq:fidelity}
F(\rho,\sigma)=[\rm{Tr}(\sqrt{\sqrt{\rho} \sigma \sqrt{\rho}})]^{2}
\end{equation}
is the fidelity \cite{NC}.
 
\begin{figure}[!t]
 	\centering	
	{\includegraphics[width=\linewidth]{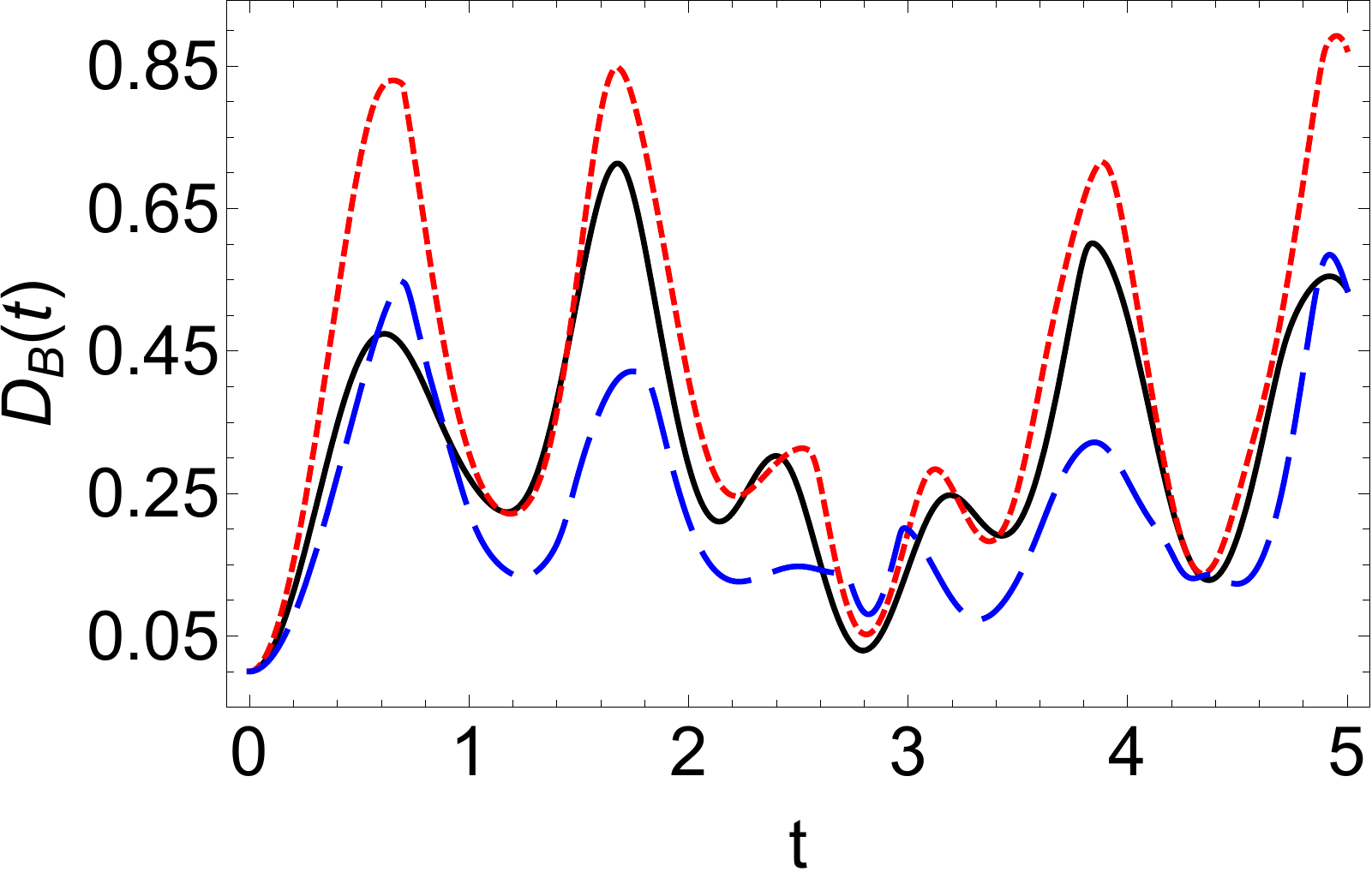}}
	\caption{\label{fig:Bures-CTQW} Time dependence of the Bures distance between the diagonal elements of the forward and backward evolved density matrices $\rho(t)$ and $\rho(-t)$, for a CTQW on a triangular chain. The curves are for the phases $\phi$ in the initial state $\ket{\psi(0)}= (\ket{1}-\exp(\text{i}\phi)\ket{2})/\sqrt{2}$ with \(\phi = \pm \pi \slash 3 \) (solid, black), \(\phi = \pm \pi \slash2 \) (dotted, red) and  \(\phi = \pm 3 \pi \slash 4 \) (dashed blue). The Bures distance for the phases \(\phi = 0 \) and \(\phi=\pm \pi\) are zero.}
 \end{figure}

In Fig. \ref{fig:Bures-CTQW}, we only use the diagonal elements of $\rho(t)$ and $\rho(-t)$ while calculating the Bures distance \(D_B(t)\) for different \(\phi\)'s. As the probability information is maintained on the diagonal elements of the density matrices, the off-diagonal elements are discarded to demonstrate the broken PTS conditions more clearly. For the phases \(0\) and  \(\pm \pi\) the PTS is not broken and the Bures distance is zero.  The largest bias in forward and backward time evolution is found for \(\phi=\pi \slash 2\),
which is different from the critical phase \(\phi=3 \pi \slash 4\) for optimum population
and entanglement transfer in CTQW over a triangular chain. We conclude that the CTQW exploits the path interference for efficient state transfer, and a certain phase
difference in the initial superposition state, quantum walk with broken PTS is possible, similar to CQW. While the chiral character of CTQW is limited to certain initial states, this can still be practically significant when the implementation of CQW is a
challenging and chiral transfer of arbitrary entangled states is not required.

%%%%%%%%%%%%%%%%%%%%%%%%%%%%%%%%% FIGURE 4 %%%%%%%%%%%%%%%%%%%%%%%%%%%%%%%%%%%%%%%%
%%%%%%%%%%%%%%%%%%%%%%%%%%%%%%%%%%%%%%%%%%%%%%%%%%%%%%%%%%%%%%%%%%%%%%%%%%%%%%%%%%%

%%%%%%%%%%%%%%%%%%%%%%%%%%%%%%%%%%%%%%%%%%%%%%%%%%%%%%%%%%%%%%%%%%%%%%%
\subsection{PTS breaking and entanglement transfer in CQW on a triangular chain and comparison with CTQW}
\label{Sec:PTSB-CQW}
%%%%%%%%%%%%%%%%%%%%%%%%%%%%%%%%%%%%%%%%%%%%%%%%%%%%%%%%%%%%%%%%%%%%%%%

%%%%%%%%%%%%%%%%%%%%%%%%%%%%%%%%% FIGURE 5 %%%%%%%%%%%%%%%%%%%%%%%%%%%%%%%%%%%%%%%%

\begin{figure}
	\centering
	{\includegraphics[width=\linewidth]{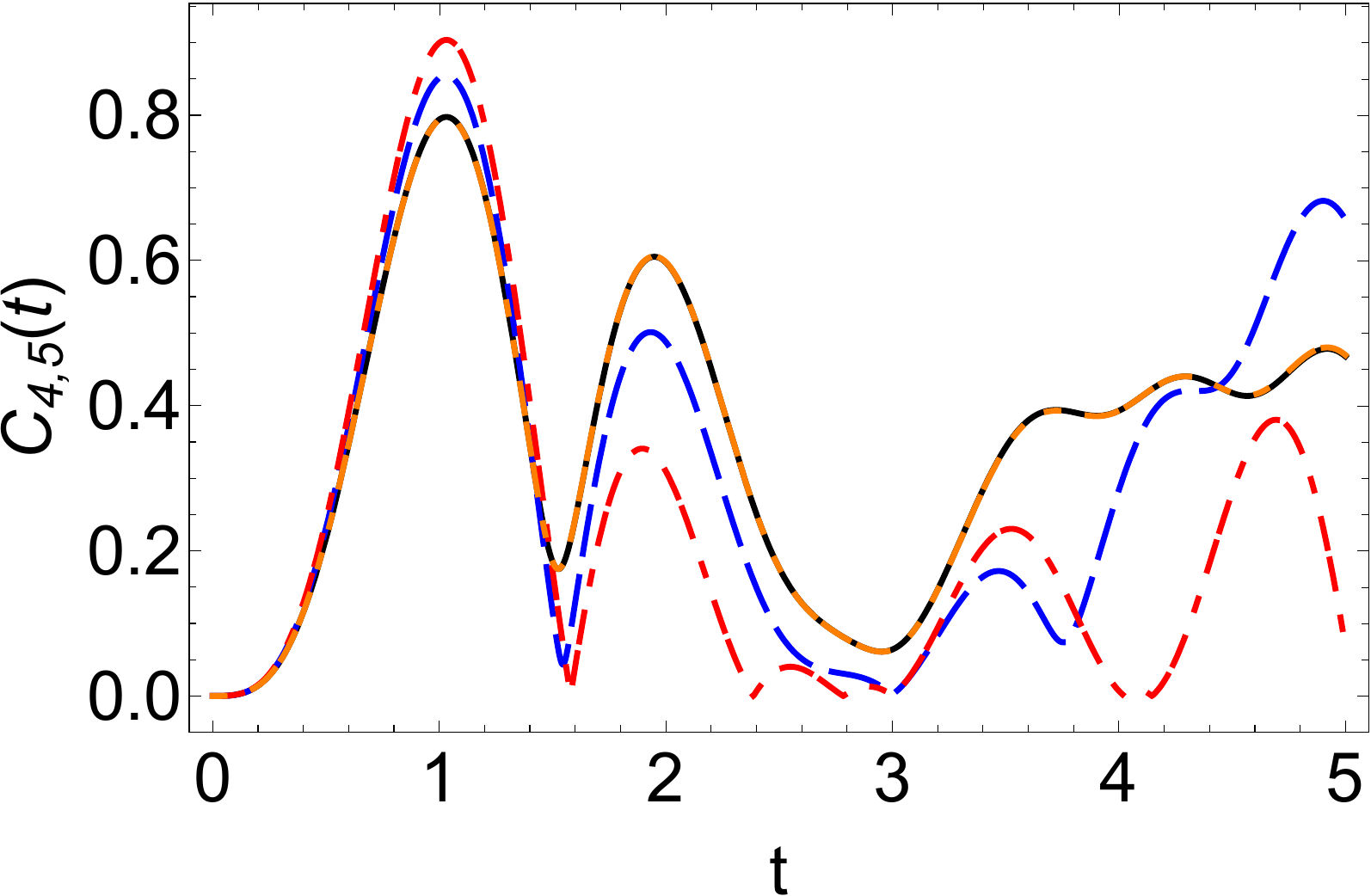}}
	\caption{\label{fig:CONC-CQW}  Time dependence of the concurrence \(C_{4,5}\) measuring the pairwise entanglement between the sites \(\ket{4}\) and \(\ket{5}\) of a triangular chain of $N=5$ sites over which an initial maximally entangled Bell state \((\ket{1}+\ket{2})\slash \sqrt{2}\) of the sites $1$ and $2$ undergoes CQW. Different curves are for the  different complex hopping coefficients of the CQW with the phases \(\theta = \pi \slash 4\) (solid, black), \(\theta = \pm \pi \slash 3\) (dashed, blue), \(\theta = \pm \pi \slash 2\) (dot-dashed-dashed, red), \(\theta = \pm 3 \pi \slash 4\) (dot-dashed, orange); same as \(\theta = \pi \slash 4\).}
\end{figure}

%%%%%%%%%%%%%%%%%%%%%%%%%%%%%%%%%%%%%%%%%%%%%%%%%%%%%%%%%%%%%%%%%%%%%%%%%%%%%%%%%

For CQW, we plot the concurrence in Fig.~\ref{fig:CONC-CQW} by using an initial Bell state with $\phi=\pi$ in Eq.~(\ref{eq:ctqw-phaseInitialState}) and different \(\theta\) in Eq.~(\ref{eq:Hcqw}). We see that the concurrence is largest for \(\theta=\pi\slash 2\) with a value \(C_{4,5}(1.02)\sim0.9\), indicating that CQW has a slight time advantage (\(\Delta t\sim0.1\)) along with a significantly higher quality transfer of entanglement compared to CTQW (cf.~Fig.~\ref{fig:CONC-CTQW}). Without plotting, we state here that a similar
conclusion applies to occupation probabilities, too. We found that CQW with \(\theta=\pi \slash 2\) yields near perfect ($P_5\sim 0.95$) state transfer $\ket{1}\rightarrow\ket{5}$ at $t\sim 1.64$.
 
We calculated the concurrences for \(\phi=\pi \slash 4\), \(\phi=\pi \slash 3\), \(\phi=\pi \slash 2\), \(\phi=3 \pi \slash 4\) and looked for the optimum \(\theta\) values. We have found that for the  state (\(\phi=\pi\)) initial Bell state, \(\theta=\pi \slash 2\) gives the optimum (maximum) concurrence \(C_{4,5}(1.02)\sim0.9\). 	
	
\begin{figure}[!t]
	\centering	
	{\includegraphics[width=\linewidth]{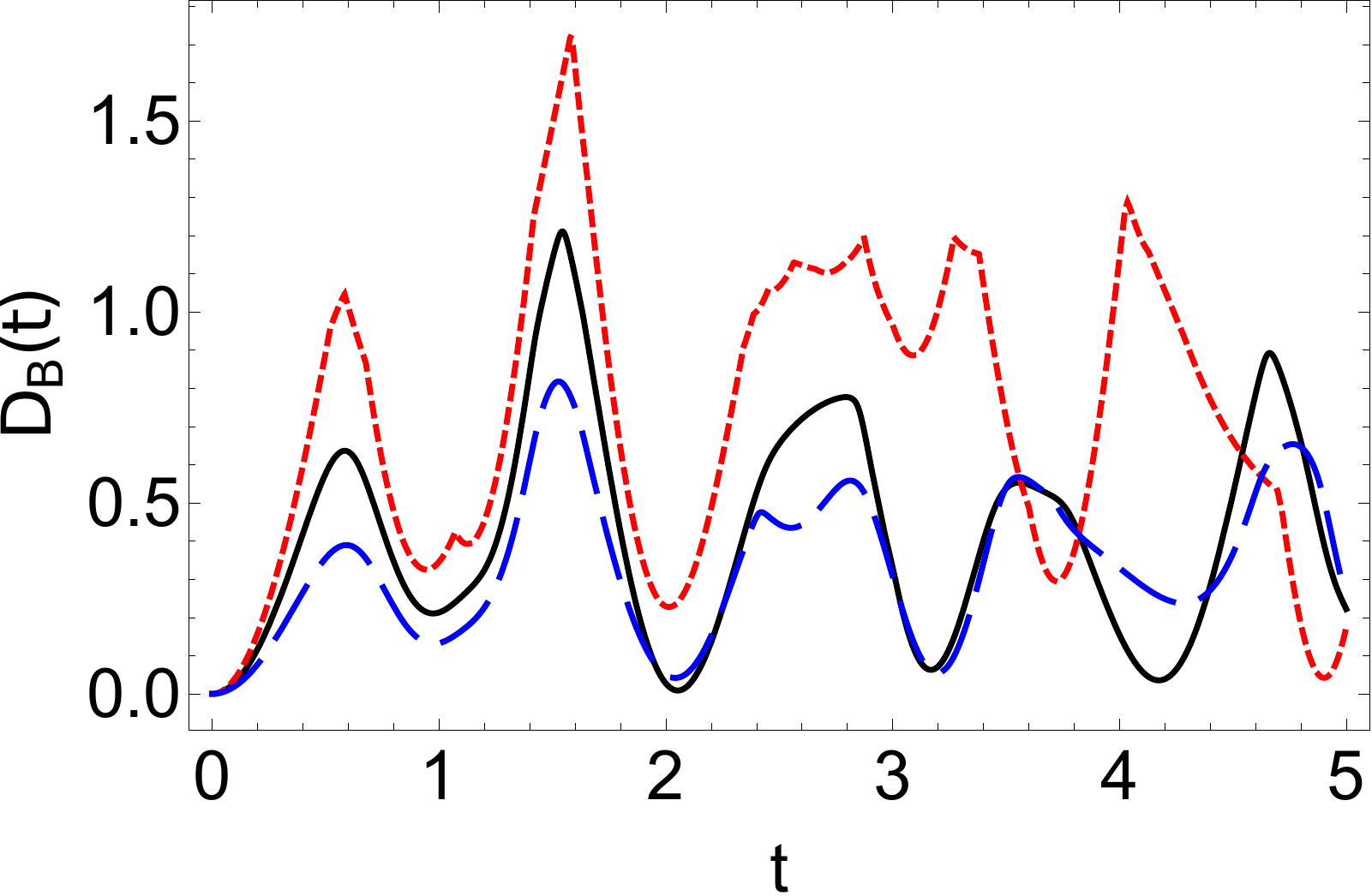}}	
	\caption{\label{fig:Bures-CQW} Time dependence of the Bures distance \(D_B(t)\) between the diagonal elements of the forward and backward evolved density matrices $\rho(t)$ and $\rho(-t)$, respectively, for a CQW on a triangular chain of $N=5$ sites. Initially the quantum walker is in a maximally entangled Bell state \((\ket{1}+\ket{2})\slash \sqrt{2}\). Different curves are for chains with different complex hopping coefficients of phases   \(\theta = \pm  \pi \slash 4 \)  (dashed, blue), \(\theta = \pm \pi \slash 3 \) (solid, black) and \(\theta = \pm \pi \slash 2 \) (dotted, red). The Bures distance for the phases \(\theta= 0 \) and \(\pm \pi\) are zero; and for \(\theta=\pi \slash 4\), \(D_B(t)\) is the same as that of \(\theta=3\pi \slash 4\).}	
\end{figure}

We plot the Bures distance \(D_B(t)\) for the diagonal elements of $\rho(t)$ and  $\rho(-t)$ in Fig.~\ref{fig:Bures-CQW}, which shows that \(D_B(t)\) is maximum for \(\theta=\pm \pi \slash 2\) (dotted, red). Remarkably,
the maximum broken PTS in the CTQW is found for $\phi=\pi/2$. This suggests that \(\phi=\theta=\pm \pi \slash 2\) is an optimal choice for the broken-PTS condition both for CTQW and CQW over
a triangular chain. The critical angle of maximum time-reversal asymmetry however coincides
with a critical angle of optimum state transfer only for the CQW. When the numerical values of \(D_B(t)\) for CTQW in Fig.~\ref{fig:Bures-CTQW} and CQW in Fig.~\ref{fig:Bures-CQW} compared, \(D_B(t)\) for CQW is numerically larger than CTQW, suggesting a larger broken PTS condition.

%%%%%%%%%%%%%%%%%%%%%%%%%%%%%% FIGURE 6 %%%%%%%%%%%%%%%%%%%%%%%%%%%%%%%%

%%%%%%%%%%%%%%%%%%%%%%%%%%%%%%%%%%%%%%%%%%%%%%%%%%%%%%%%%%%%%%%%%%%%%%%%%%%%%%%%%

\begin{figure}
  	\centering
  	\includegraphics[width=\linewidth]{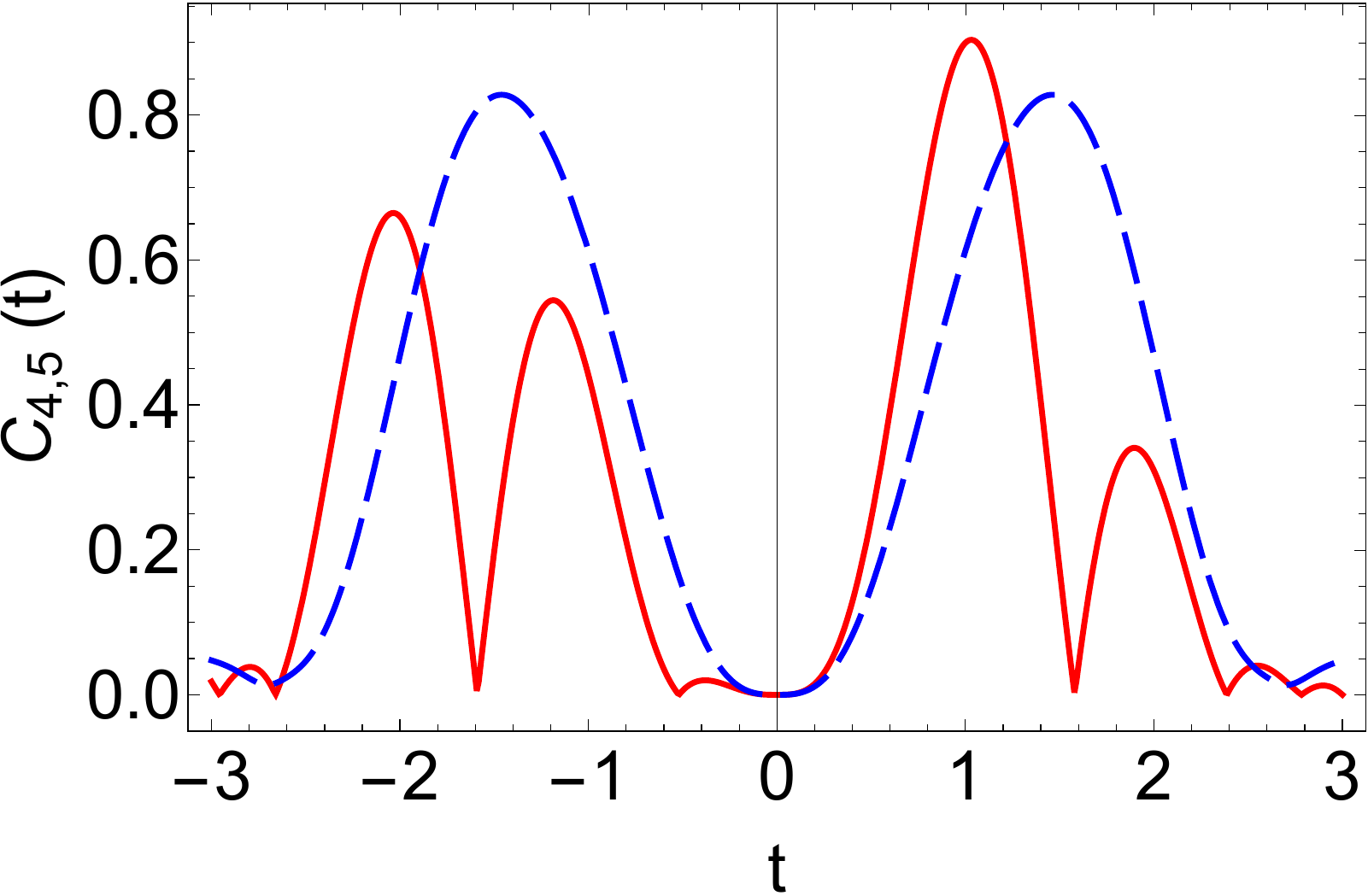}
 	\caption{\label{fig:concCQWCTQW} Time($t$) dependence of the concurrence \(C_{4,5}(t)\) to measure the entanglement between the sites \(\ket{4}\) and \(\ket{5}\) for an initially maximally entangled Bell state \((\ket{1}+\ket{2})\slash \sqrt{2}\) of the sites \(\ket{1}\) and \(\ket{2}\) of a particle that makes CTQW (dashed, blue) and CQW (solid, red) with \(\theta=\pi \slash 2\) on a triangular chain of $N=5$ sites. We take the phases of the complex hopping coefficients as \(\theta=\pi \slash 2\) for CQW;
 	while for CTQW hopping coefficients are real with \(\theta=0\).}
\end{figure}

In Fig.~\ref{fig:concCQWCTQW}, the concurrences for CQW and CTQW with an initial state of \((\ket{1}+\ket{2})\slash \sqrt{2}\) are plotted. The solid red curve represents the CQW case and the dashed blue line is for the CTQW case. This plot also demonstrates the broken PTS in the CQW case. Here, one can notice the relatively higher entanglement transfer quality in CQW. In addition, the transfer time is shorter in the case of CQW by $\Delta t \sim 0.4$ .

\begin{figure}[t!]
	\centering
	\includegraphics[scale=0.6 ,fbox]{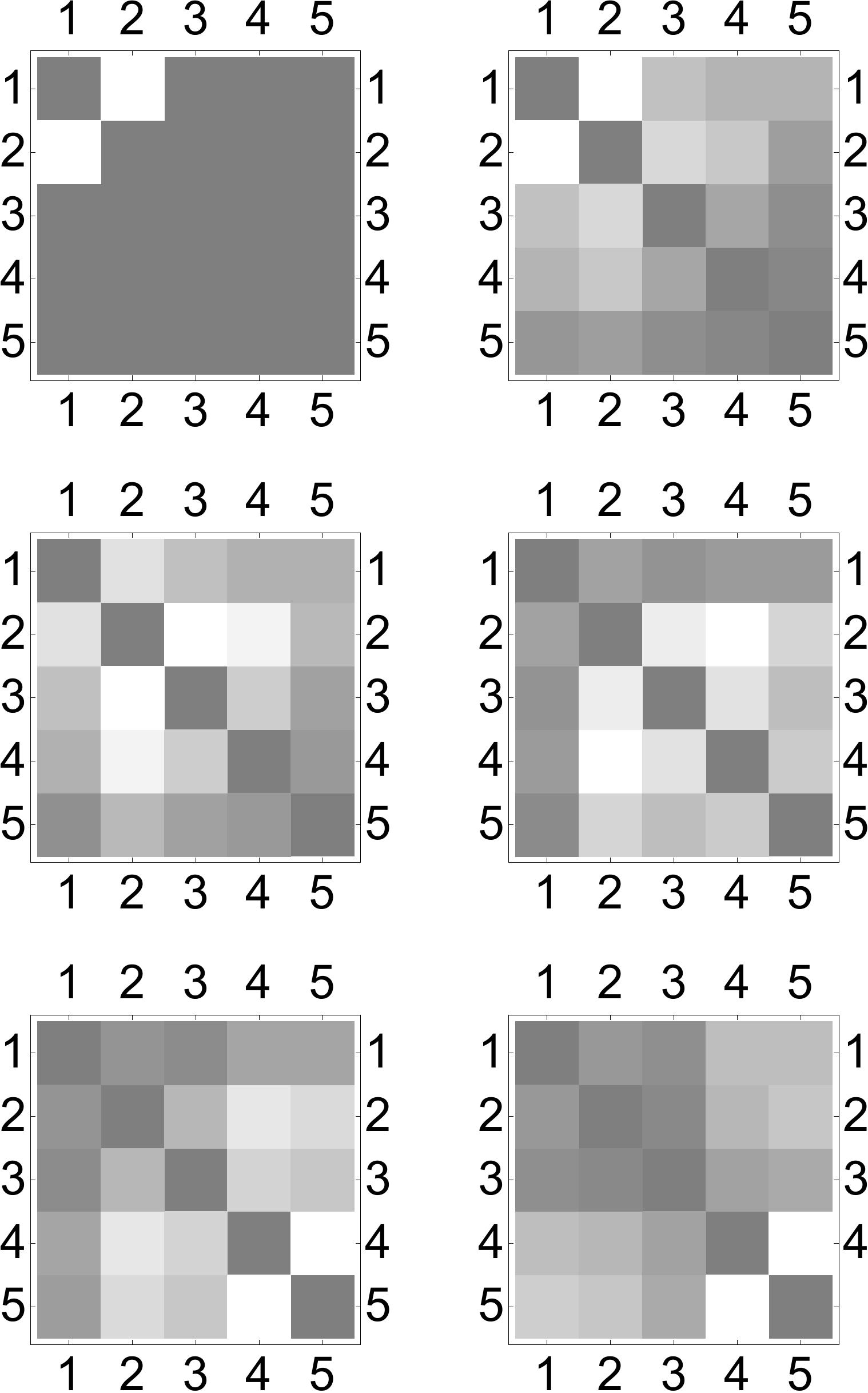}
	\qquad
	\caption{\label{fig:walkEntangled}The time  evolution of concurrence $C_{i,j}$, measuring the pairwise entanglement between the sites $|i\rangle$ and $|j\rangle$, is shown as a matrix with $i,j=1..5$. Each square in these matrices stands for the value of the concurrence $C_{i,j}$. Colors from light to dark scale with $1$ to $0$, respectively. Initially (at $t=0$), the quantum walker is injected in the maximally entangled Bell state of the sites $1$ and $2$ with $C_{1,2}=1$ (upper left panel) to undergo CQW with complex hopping coefficients with phase $\theta=\pi/2$. As time progresses, one can notice the unidirectional transfer of entanglement (light colored pair of squares) to the rightmost sites $4$ and $5$. The panels are for the $t=0$ (upper left), $t=0.2$ (upper right), $t=0.4$ (middle left), $t=0.6$ (middle right), $t=0.8$ (bottom left), and $t=1$ (bottom right).}
\end{figure}

To demonstrate the entanglement transfer in the short-time regime, we plot the dynamics of the concurrences $C_{i,j}$ that measure entanglement between every pair of sites $(i,j)$ of the triangular chain in Fig.~\ref{fig:walkEntangled}. One can see the transfer of entanglement from the sites \((\ket{1},\ket{2})\) to \((\ket{4},\ket{5})\). Although a spread over the sites is present, entanglement propagates mainly as \((\ket{1},\ket{2})\rightarrow(\ket{2},\ket{3})\rightarrow(\ket{2},\ket{4})\rightarrow(\ket{3},\ket{4})\rightarrow(\ket{4},\ket{5})\). If the success fidelity or the concurrence is sufficient, the entanglement can be collected at the end of the chain in this short-time regime. On the other hand, after multiple scatterings between the ends of the chain, the entanglement transfer can be enhanced at the cost of longer transfer time. 

%%%%%%%%%%%%%%%%%%%%%%%%%%%%% FIGURE 7 %%%%%%%%%%%%%%%%%%%%%%%%%%%%%%%%%%%%%%%%%%%%%%
%%%%%%%%%%%%%%%%%%%%%%%%%%%%%%%%%%%%%%%%%%%%%%%%%%%%%%%%%%%%%%%%%%%%%%%

%%%%%%%%%%%%%%%%%%%%%%%%%%% FIGURE 8 %%%%%%%%%%%%%%%%%%%%%%%%%%%%%%%%
%%%%%%%%%%%%%%%%%%%%%%%%%%%%%%%%%%%%%%%%%%%%%%%%%%%%%%%%%%%%%%%%%%%%%%%

\begin{figure}[h]
	\centering
	\includegraphics[scale=0.8]{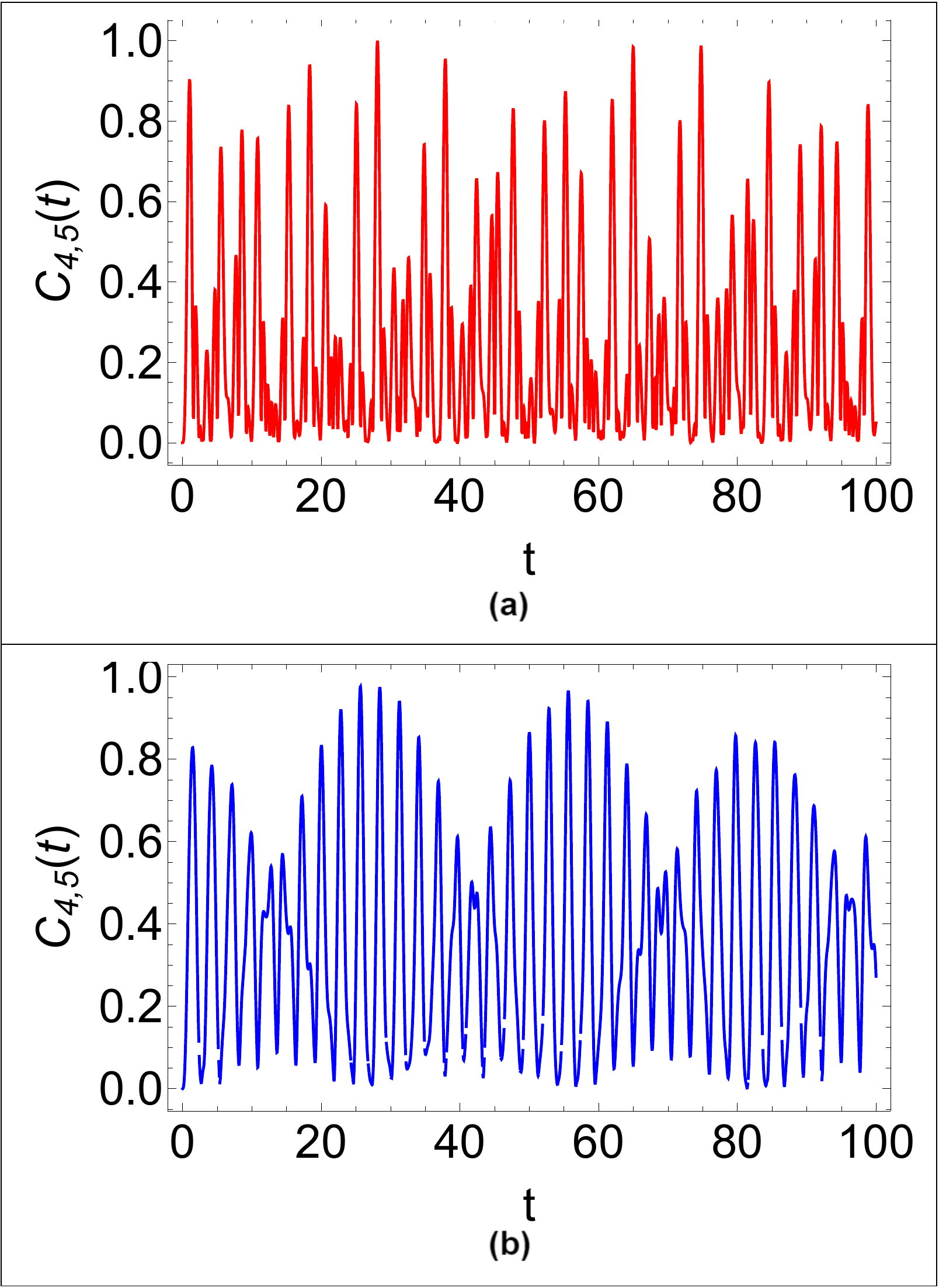}
	\qquad
	\caption{\label{fig:figBackScatter}Time dependence of the concurrence $C_{4,5}(t)$ of CQW (solid blue curve) and CTQW (solid red curve) for the long-time evolution $(t = 0 − 100)$. In this plot, we have considered the initial state $\ket{\psi_\text{spatial}(0)}=(\ket{1}+\ket{2})/\sqrt{2}$.}
\end{figure}

In Fig.~\ref{fig:figBackScatter}, we plot the long-time behavior of the process for both CTQW and CQW. CQW demonstrates a higher concurrence peak in the short-time regime for the initial Bell state ($\phi=0$). 
When the fidelities are considered, the longer-time entanglement transfer fidelity is higher than
the short-time regime's fidelities for CQW and CTQW. Both allow for PGST of the entanglement with $C_{4,5} = 0.999$ at
$t = 28.1$ and
with a concurrence of
$C_{4,5} = 0.971$ at $t = 25.7$, respectively. 
These observations depend on the initial state and the chain size.
Though not shown here, we numerically verified that CTQW gives a higher concurrence than CQW for certain initial conditions in the short-time regime  (e.g., for $\phi=\pi/2$). 
Hence, we conclude that breaking PTS either by CQW for any initial condition or by CTQW for certain initial conditions gives comparable and high entanglement transfer performance in a short-time regime, which can be further enhanced to PGST in a long-time regime. 
The successful entanglement transfer (with a concurrence of more than $0.9$) is limited to chains shorter than $N\sim 9$ sites, as discussed in Sec.~\ref{Sec:ChainSize}.

%%%%%%%%%%%%%%%%%%%%%%%%%%% FIGURE 8 %%%%%%%%%%%%%%%%%%%%%%%%%%%%%%%%
%%%%%%%%%%%%%%%%%%%%%%%%%%%%%%%%%%%%%%%%%%%%%%%%%%%%%%%%%%%%%%%%%%%%%%%

%----------------------------------------------------------------
\subsection{Transfer of mixed Werner-States on the triangular chain}
\label{Sec:WernerState}
%----------------------------------------------------------------
Having demonstrated the role of pure entangled states under the CQW scheme, it is natural to investigate the behavior of Werner-type mixed states under CQW with \(\theta=\pi \slash 2\) phase \cite{werner}. We introduce the Werner-like state
\begin{equation}\label{eq:wernerState}
\rho_{\text{Werner}}(b) = b \rho(0) + (1-b)\rho_{\text{mixed}},
\end{equation}
where $\rho_{\text{mixed}}$ is the maximally mixed state within the manifold of injection site states $\ket{1}$ and $\ket{2}$.
\begin{equation}\label{eq:rhoMixed}
\rho_{\text{mixed}}=\frac{1}{2}\sum_{i=1}^{2}\ket{i}\bra{i}.
\end{equation}
We use the maximally entangled state within the manifold of injection site states, \(\ket{\psi(0)}=(\ket{1}+\ket{2})/\sqrt{2}\) to define the initial state density matrix \(\rho(0)=\ket{\psi_\text{spatial}(0)}\bra{\psi_\text{spatial}(0)}\).

To investigate the behaviour of entanglement transfer with respect to time, we calculate the fidelity $F(\rho(t),\rho_{\text{target}})$ as in Eq.~(\ref{eq:fidelity}) with the matrices $\rho(t)$ and $\rho_{\text{target}}$. Here, $\rho_{\rm{target}}$ represents the desired ideally transferred state
\begin{equation}\rho_{\text{target}}=\frac{1}{2}
\begin{bmatrix}
0&0&0&0&0\\
0&0&0&0&0\\
0&0&0&0&0\\
0&0&0&1&b\\
0&0&0&b&1\\
\end{bmatrix}.
\end{equation}

\begin{figure}[t!]
	\centering
	\includegraphics[width=\linewidth]{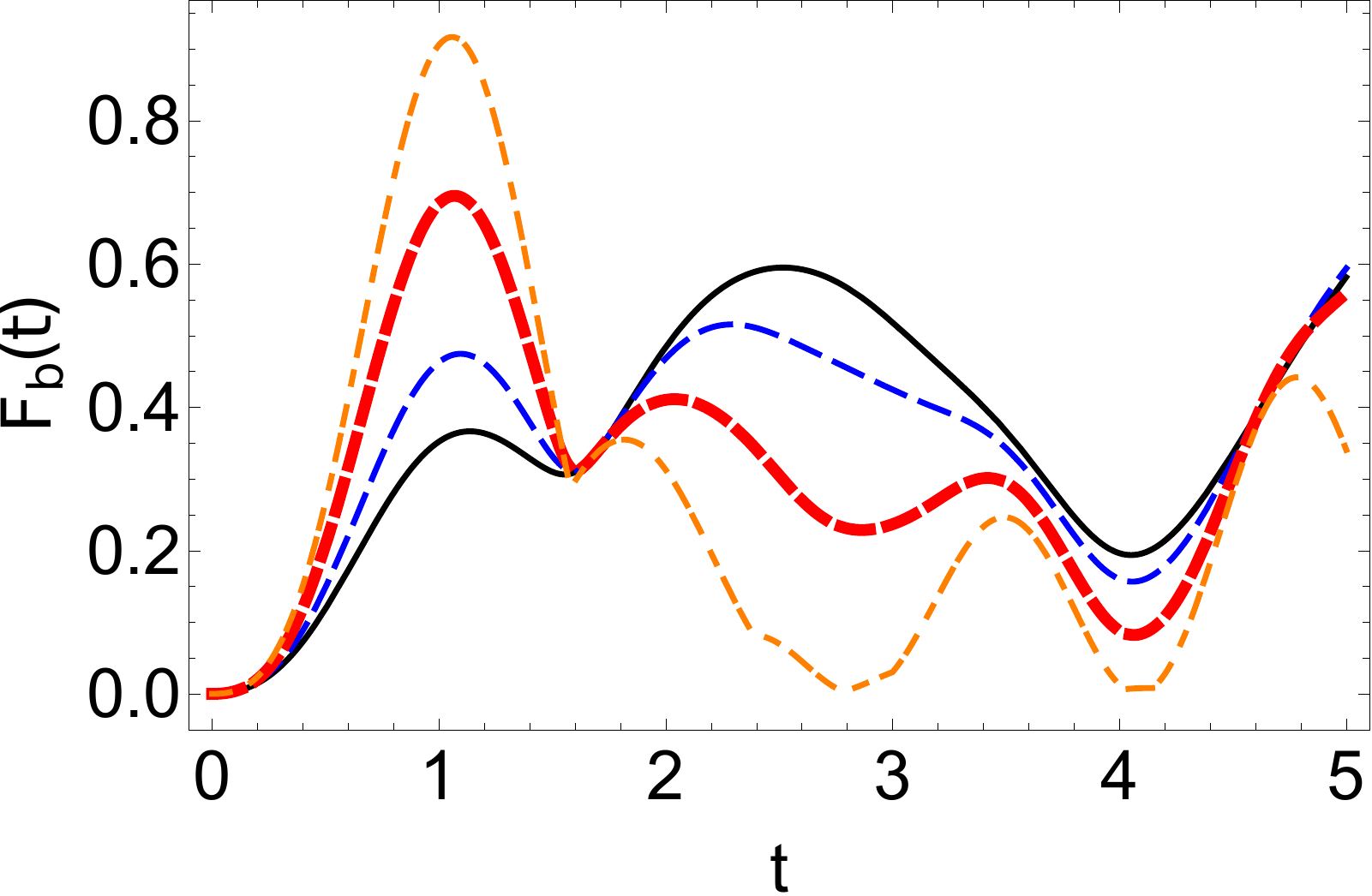}
	\caption{\label{fig:wernerFig}Time dependence of the Fidelity  $F(\rho(t),\rho_{\text{target}})$ between \(\rho(t)\) and  \(\rho_{\text{Werner}}\) as a function of time $t$ with an initial Werner state under the scheme of CQW on a triangular chain of $N=5$ sites with complex edge weights of phase $\theta=\pi/2$. Where \(b=-1\slash 4\) as solid black, \(b=0\) as dashed blue, \(b=1\slash 2\) as thick dashed red and finally, maximally entangled bell state \(b=1\) is dot-dashed orange.}
\end{figure}

In Fig.~\ref{fig:wernerFig}, we plot the time behavior of the Fidelity,  $F(\rho(t),\rho_{\text{target}})$, for different \(b\) values. Clearly, the pure maximally entangled state \(\rho_{\text{Werner}}(b=1)=\rho(t=0)\) yields the best entanglement transfer. On the contrary, fidelities closer to zero is observed at the maximally mixed state \(\rho_{\text{Werner}}(b=0)=\rho_{\text{mixed}}\).

%%%%%%%%%%%%%%%%%%%%%%%%%%%%%%%%%%%%%%%%%%%%%%%%%%%%%%%%%%%%%%%%%
%%%%%%%%%%%%%%%%%%%%%%%%%%%%%%%%%%%%%%%%%%%%%%%%%%%%%%%%%%%%%%%%%

%----------------------------------------------------------------
\subsection{Scaling of entanglement transfer quality and time with respect to the chain size }
\label{Sec:ChainSize}
%----------------------------------------------------------------

\begin{figure}[!t]
	\centering
	\subfloat[\label{fig:tmax}]{\includegraphics[scale=0.50]
		{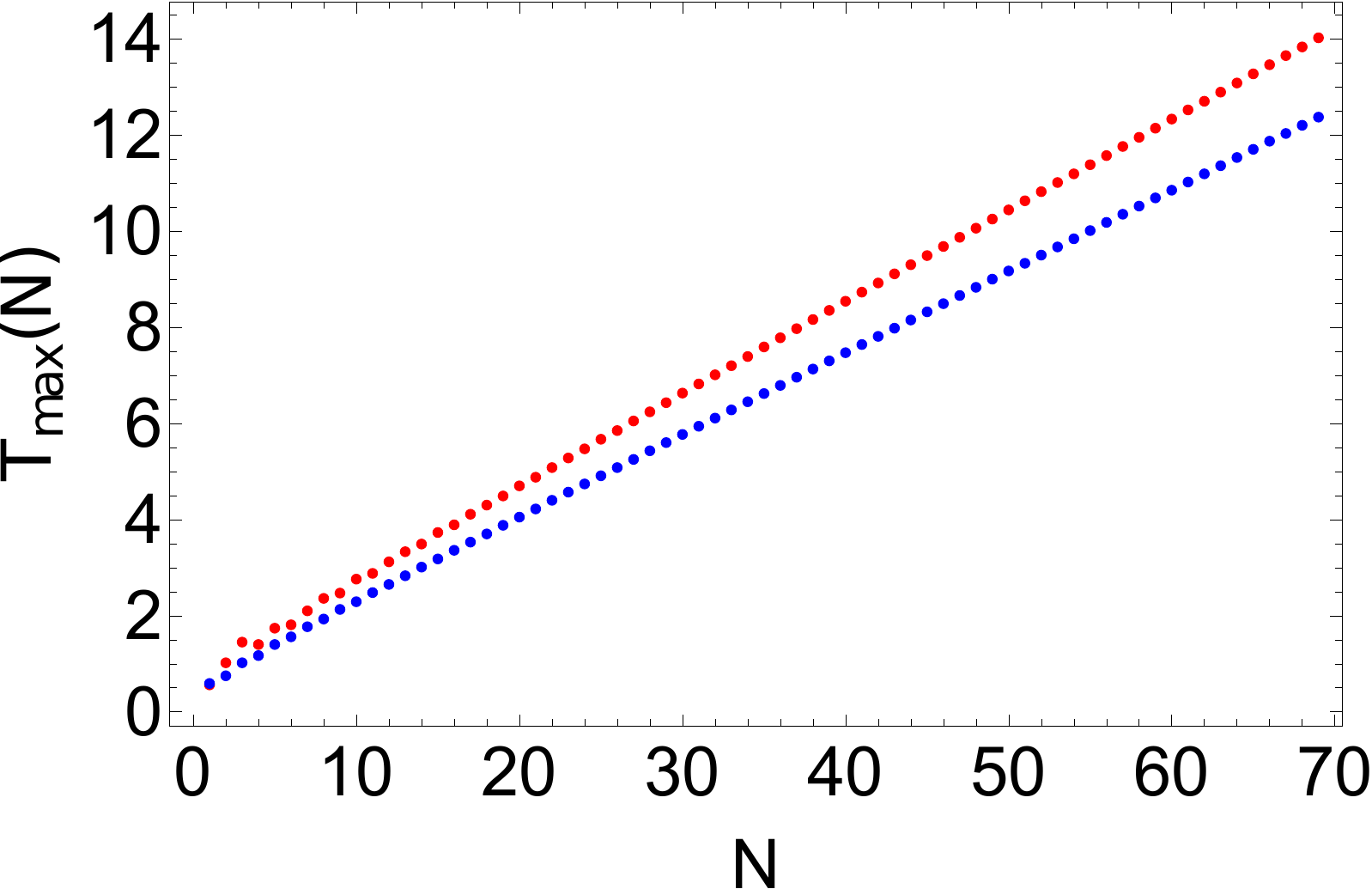}}
    \qquad
	\subfloat[\label{fig:tmaxconc}]{\includegraphics[scale=0.50]
		{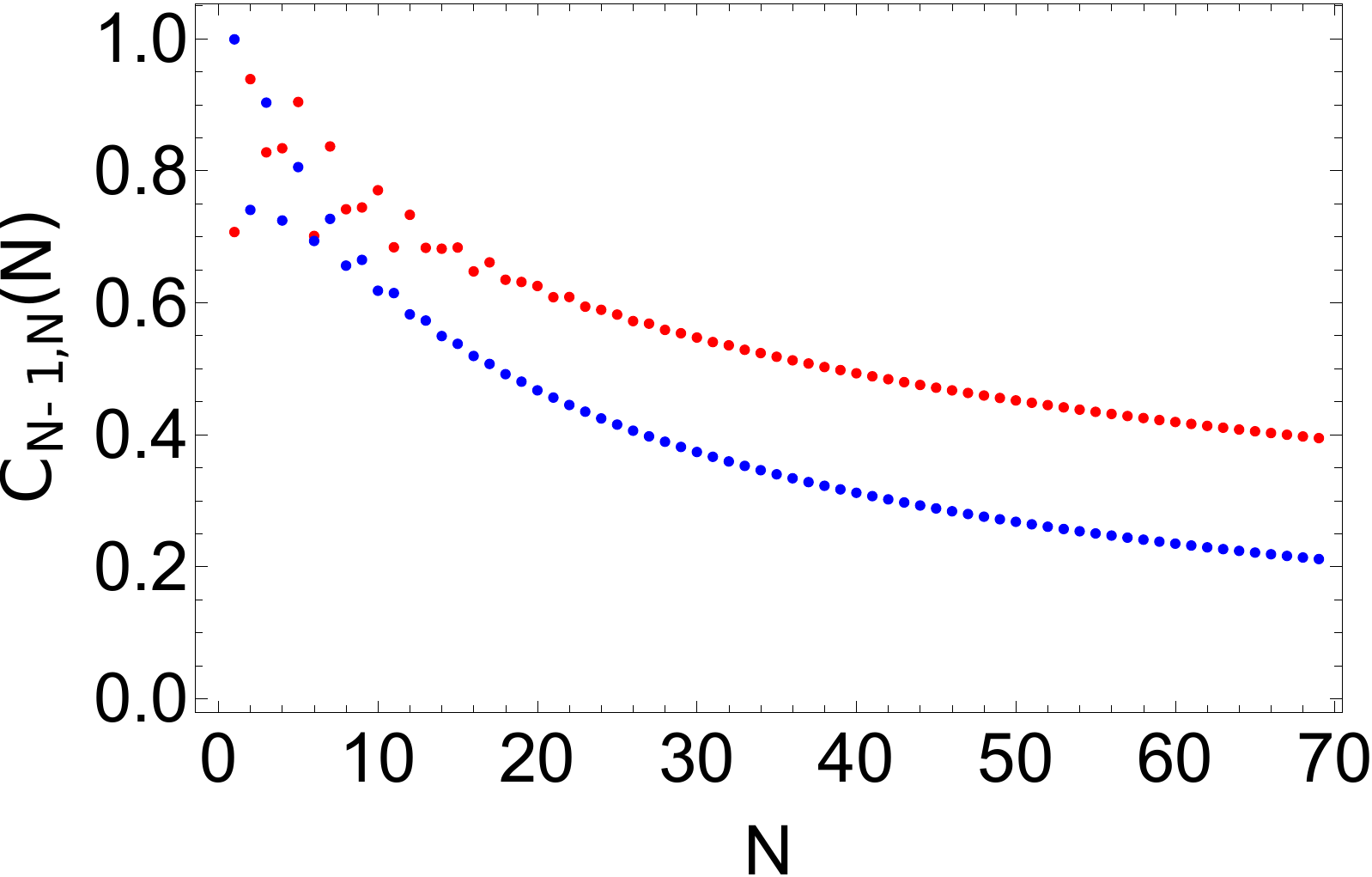}}
	\caption{\label{fig:tmaxc}\textbf{(a)} Scaling of state transfer time with respect to the chain size. Red dots represent $\theta=0$ and blue dots represent $\theta=\pi/2$~\textbf{(b)} Concurrence value at the state transfer time. Red dots represent $\theta=0$ and blue dots represent $\theta=\pi/2$}		
\end{figure}

Until now, entanglement transfer on a chain of $N=5$ sites has been discussed. In this subsection, we first investigate the entanglement transfer using CQW in the short-time regime by calculating the time $T_\text{max}$ and value $C_{N-1,N}$ of the first peak of the concurrence for chains of up to $N=71$ sites. Fig.~\ref{fig:tmax} shows that $T_\text{max}$ (which we refer to as transfer time) scales linearly with $N$, consistent with previous works~\cite{tony1,tony2, tony3}. Fig.~\ref{fig:tmaxconc} shows that the entanglement transfer quality decreases severely in chains longer than $N\sim 9$ sites.
The figures include two different chiral phases $\theta$ values and indicate similar behavior. 

Next, we explore the long-time entanglement transfer dynamics by assuming a waiting time of
$t=500$. In this case, we determine and fix an ideal value of $\theta$ for given $N$ and initial state to find the maximum concurrence, which is not necessarily the first peak. The optimum $\theta$ for the initial state with $\phi=\pi$ is found to be $\pm \pi/2$, whose sign depends on $N$. The results are given in Table~\ref{tab:cqw},
which shows the maximum concurrence and when it occurs for given $N$ and the corresponding optimum $\theta$. For the same initial condition, the results for CTQW are presented in 
Table~\ref{tab:ctqw}. While the tables report the results for the chains with up to $N=33$ sites,
the concurrence reduces to very low values after $N\sim 9$ sites. We can see that both
CQW and CTQW methods to break PTS yield highly successful transfer of entangled states for
relatively small graphs ($N<9$). For such graphs ( ($N<9$), CQW is faster than CTQW to transfer the entanglement, and its success is slightly higher (also cf.~Fig.~\ref{fig:concCQWCTQW} where the same results are found for another initial condition ($\phi=0$) and in the short-time regime).

%%%%%%%%%%%%%%%%%%%%%%%%%%%%% FIGURE 13 %%%%%%%%%%%%%%%%%%%%%%%%%%%%%%
%%%%%%%%%%%%%%%%%%%%%%%%%%%%%%%%%%%%%%%%%%%%%%%%%%%%%%%%%%%%%%%%%%%

\begin{table}
\centering
\begin{tabular}{||c | c c c  || } 
 \hline
 Chain Size $N$ & Time $(t)$ & Concurrence & Optimum Chiral Phase $(\theta)$   \\ [0.5ex] 
 \hline\hline
 5 & 55.4 & 0.999 & $-\pi/ 2$  \\ 
 \hline
 7 & 85.1 & 0.992 & $\pi/ 2$  \\
 \hline
 9 & 2.9 &  0.947 & $-\pi/ 2$ \\
 \hline
 11 & 321.3 & 0.900 & $-\pi/ 2$ \\
 \hline
 13 & 397.6 & 0.885 & $-\pi/ 2$ \\ 
 \hline
 15 & 4.5 & 0.874 & $-\pi/ 2$ \\ 
 \hline
 17 & 136.2 & 0.700 & $-\pi/ 2$ \\ 
 \hline
 19 & 68.6 & 0.714 & $-\pi/ 2$ \\ 
 \hline
 21 & 6.1 & 0.814 & $-\pi/ 2$ \\ 
 \hline
  23 & 416 & 0.635 & $-\pi/ 2$ \\ 
 \hline
  25 & 88.5 & 0.711 & $-\pi/ 2$ \\ 
 \hline
  27 & 7.7 & 0.764 & $-\pi/ 2$ \\ 
 \hline
  29 & 125.8 & 0.593 & $\pi/ 2$  \\ 
 \hline
  31 & 376.5 & 0.736 & $\pi/ 2$  \\ 
 \hline
  33 & 9.3 & 0.718 & $-\pi/ 2$ \\ [1ex] 
 \hline
\end{tabular}
\caption{\label{tab:cqw} The maximum concurrence and the transfer time table for the longer-time CQW scenario $(t=500)$ with the initial state phase $\phi=\pi$ along with the optimal phases for these parameters.}
\end{table}

\begin{table}
\centering
\begin{tabular}{||c | c c  ||} 
 \hline
 Chain Size $N$ & Time $(t)$ & Concurrence\\ 
 [0.5ex] 
 \hline\hline
 5 & 193.9 & 0.993 \\ 
 \hline
 7 & 342.8 & 0.979  \\
 \hline
 9 & 410.6 &  0.900\\
 \hline
 11 &  482.7 & 0.805\\
 \hline
 13 & 498.3 & 0.749\\
 \hline
 15 & 288.2 & 0.748  \\  
 \hline
 17 & 82.5 & 0.697 \\  
 \hline
 19 & 4.1 & 0.661 \\ 
 \hline
 21 & 4.5 & 0.631 \\ 
 \hline
  23 & 4.9 & 0.608  \\ 
 \hline
  25 & 5.3 & 0.594 \\ 
 \hline
  27 & 5.7 & 0.581 \\
 \hline
  29 & 6.1 & 0.567 \\ 
 \hline
  31 &  6.4 & 0.552   \\
 \hline
  33 & 6.8 & 0.540 \\ [1ex] 
 \hline
\end{tabular}
\caption{\label{tab:ctqw} The maximum concurrence and the transfer time table for the longer-time CTQW scenario $(t=500)$ with the initial state phase $\phi=\pi$.}
\end{table}

%%%%%%%%%%%%%%%%%%%%%%%%%%%%%%%%%%%%%%%%%%%%%%%%%%%%%%%%%%%%%%%%%

%%%%%%%%%%%%%%%%%%%%%%%%%%%%%%%%%%%%%%%%%%%%%%%%%%%%%%%%%%%%%%%%%

%%%%%%%%%%%%%%%%%%%%%%%%%%%%% FIGURE 10 %%%%%%%%%%%%%%%%%%%%%%%%%%%%%%
\begin{figure}[!t]
	\centering
	\subfloat[\label{fig:5cycleA}]{\includegraphics[scale=0.50]
		{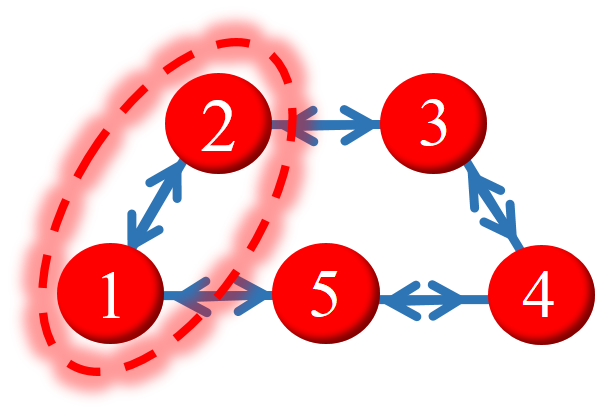}}
    \qquad
	\subfloat[\label{fig:5starB}]{\includegraphics[scale=0.50]
		{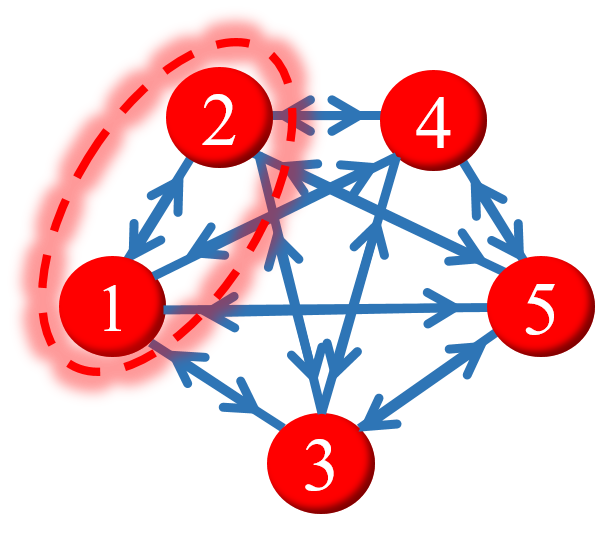}}
	\caption{\label{fig:5CycleStar}\textbf{(a)} Graph of a linear chain of 5 vertices arranged as an odd number cycle.~\textbf{(b)} Graph of a linear chain of 5 vortices arranged as a pentagram with five-pointed star-like diagonal connections. }		
\end{figure}
%%%%%%%%%%%%%%%%%%%%%%%%%%%%%%%%%%%%%%%%%%%%%%%%%%%%%%%%%%%%%%%%%%%

%%%%%%%%%%%%%%%%%%% FIGURE 11 %%%%%%%%%%%%%%%%%%%%%%%%%
\begin{figure}[!t]
	\centering
	{\includegraphics[width=\linewidth]{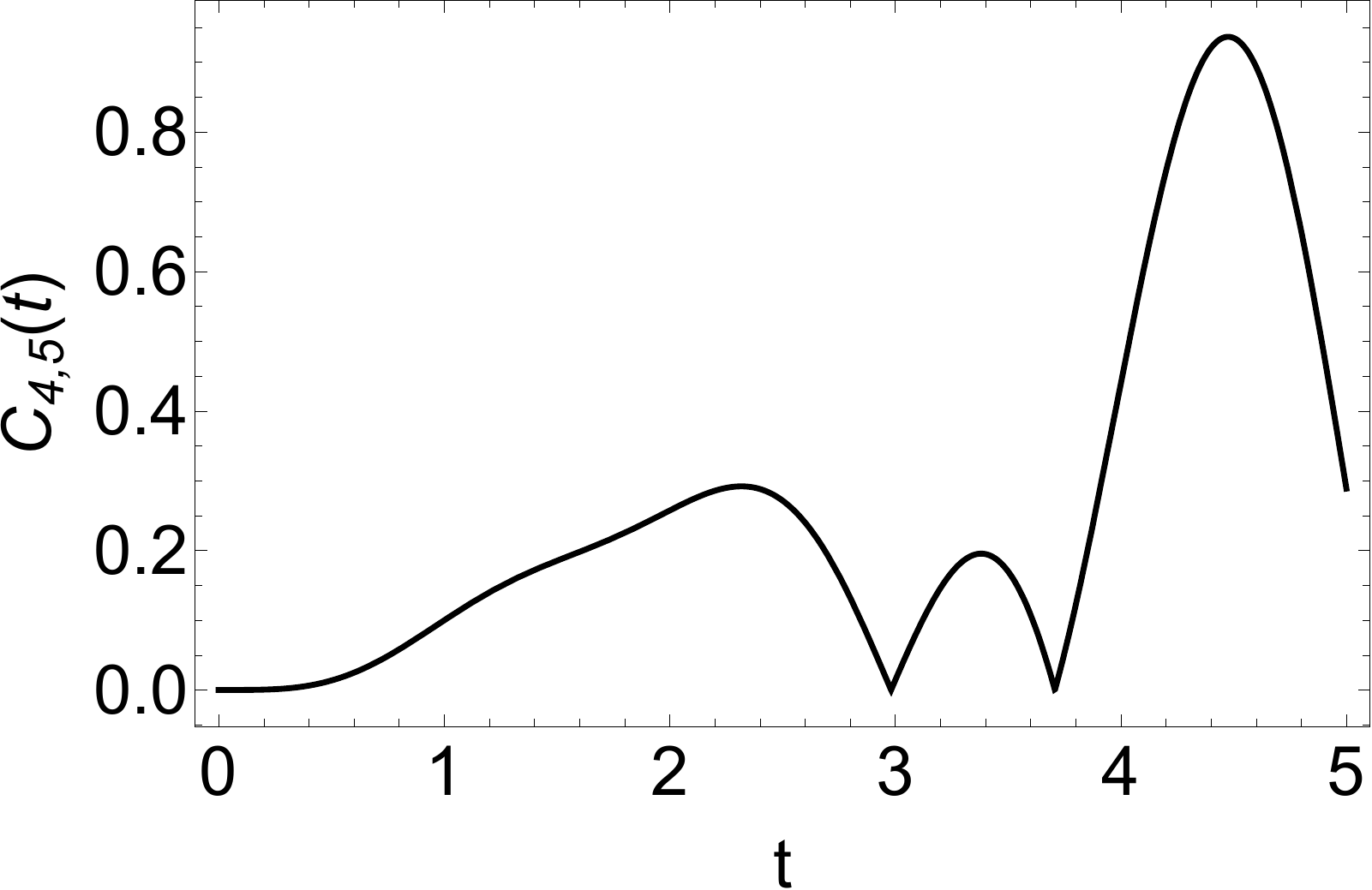}}
	\caption{\label{appFig1} Time($t$) dependence of the concurrence \(C_{4,5}(t)\) to measure the entanglement between the sites \(\ket{4}\) and \(\ket{5}\) on a \(5 \times 5\) circulant graph with only nearest-neighbour interactions for an initially maximally entangled Bell state \((\ket{1}+\ket{2})\slash \sqrt{2}\) of the sites \(\ket{1}\) and \(\ket{2}\) of a particle that makes CQW.}
\end{figure}
%%%%%%%%%%%%%%%%%%%%%%%%%%%%%%%%%%%%%%%%%%%%%%%%%%%%%%%

%%%%%%%%%%%%%%%%%%%%%%%%%%%%%%%%%%%%%%%%%%%%%%%%%%%%%%%
\begin{figure}[!t]
	\centering
	{\includegraphics[width=\linewidth]{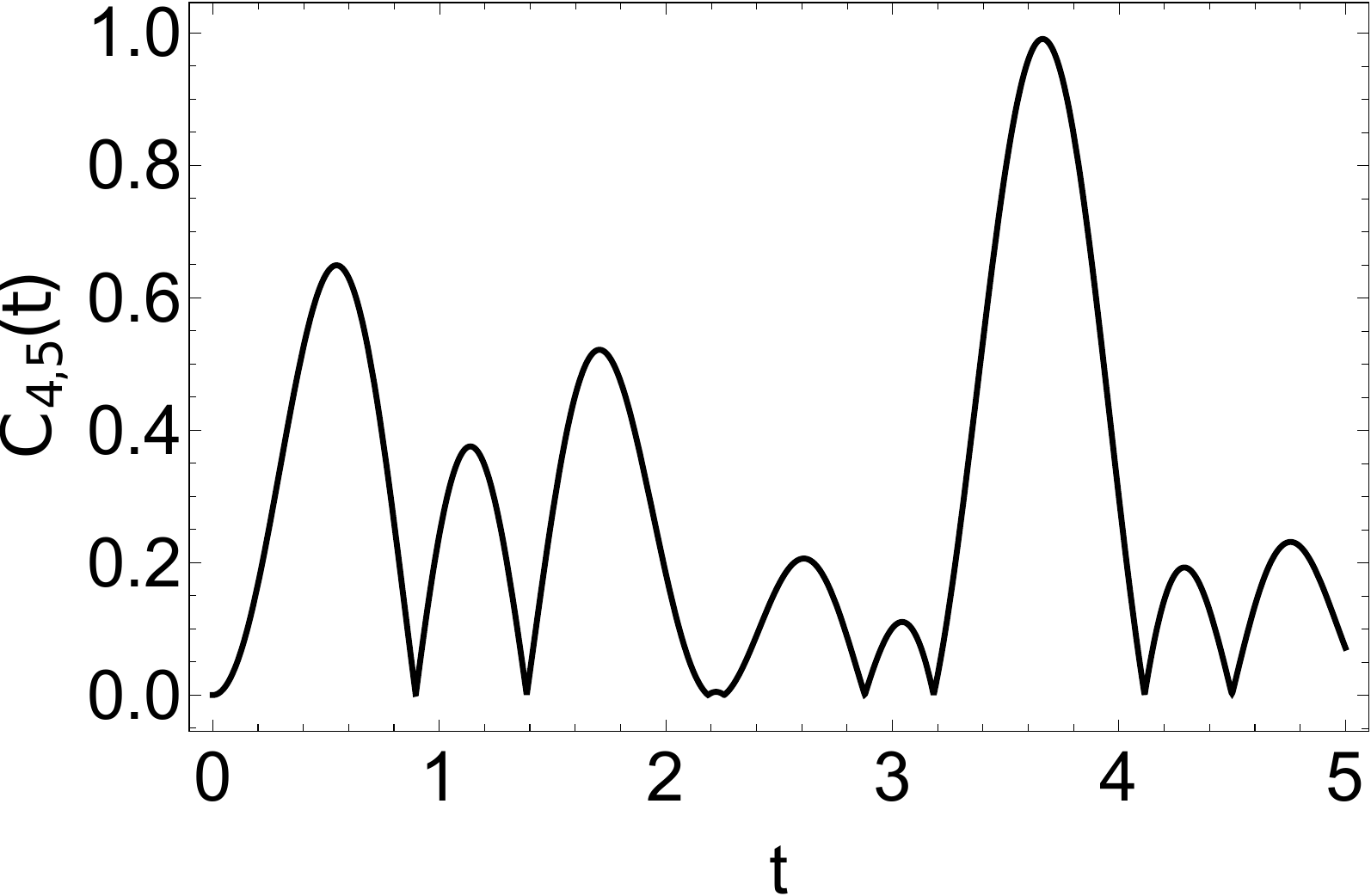}}
	\caption{\label{appFig2} Time($t$) dependence of the concurrence \(C_{4,5}(t)\) to measure the entanglement between the sites \(\ket{4}\) and \(\ket{5}\) on a complete pentagram graph for an initially maximally entangled Bell state \((\ket{1}+\ket{2})\slash \sqrt{2}\) of the sites \(\ket{1}\) and \(\ket{2}\) of a particle that makes CQW.}
\end{figure}
%%%%%%%%%%%%%%%%%%%%%%%%%%%%%%%%%%%%%%%%%%%%%%%%%%%%%%
%----------------------------------------------------------------
\section{Conclusion}
\label{Sec:Conclusion}
%----------------------------------------------------------------
We explored the transfer of spatial entanglement of a single spin excitation (which we call particle) undergoing either CQW or CTQW on a triangular chain. We found that particle transfer to the end of the chain is more successful if the particle is injected simultaneously from the leftmost pair of sites in a specific
Bell-type superposition state. The success, measured by the rightmost site's occupation probability, depends on the relative phase $\phi$ between the site states in the initial quantum superposition. Using the Bures distance between the forward and backward time evolved states, we examined the dynamics of PTS breaking at different $\phi$. We conclude that PTS breaking and the success of entangled state transfer via CTQW vary with $\phi$. We explained the physical mechanism in terms of the role of the relative phase $\phi$ in the initial state played in the path interference in the triangular chain which eventually determines the quality of state transfer via CTQW. The success and PTS breaking character of CTQW is limited to certain initial
states with strict initial phase values. 

The chiral phase angle in CQW brings additional flexibility and generality to transfer arbitrary entangled states, which is not possible with CTQW. The chiral phase can be used to optimize transfer success. Even for those entangled states that can be transferred by CTQW, using optimum chiral phases, entanglement transfer with CQW is found to be faster and more successful for small graphs with less than $9$ sites. 
In our examinations, we also considered long chains (about 70 sites). When longer triangular chains are considered, the entanglement transfer success is severely reduced for both CQW and CTQW. 

We examined both short-time and long-time dynamics of entanglement transfer. In the short-time regime, the first peak of the concurrence is used to probe the entanglement transfer.  The time when the first peak emerges (entanglement transfer time) scales linearly with the chain size, as expected from the earlier 
works~\cite{tony1, tony2, tony3}. Longer-time regime is used
to look for a global maximum in entanglement dynamics, and
hence it can give higher entanglement transfer success at the
cost of longer waiting times. Speed and success advantages
of CQW over CTQW for certain initial states remain in the
longer-time regime as well. Breaking PTS strongly either by
CQW for any initial condition or by CTQW for certain initial
conditions give comparable and high entanglement transfer
performance in a short-time regime, which can be further enhanced
in long-time regime; though short-time regime can
be more practical for real applications open to environmental
quantum decoherence effects. 

In summary, if CTQW is capable
to transfer entanglement with PTS breaking for a certain initial state, then
the performance of transfer is comparable to CQW. Hence, if implementing
CQW is challenging and transfer of arbitrary entangled states is not
required, we conclude that CTQW can be preferred over CQW. On the other hand,
if the optimum transfer of arbitrary entangled states with PTS breaking character is required than it is necessary to implement CQW. Our main conclusion is foundational in nature, based upon the physical mechanism of PTS breaking in terms of the 
path interference
and phases in the initial state and hopping coefficients, and hence, is
independent of any physical embodiment.

In addition, we explored the behavior of various mixed Werner states under our CQW scheme. We found that the purest maximally entangled state yields the best state transfer.

Our results can help to understand the interplay of PTS breaking and entanglement transfer and practically to design optimum chiral lattices for the transfer of entangled states in physical
platforms such as plasmonic non-Hermitian coupled waveguides~\cite{Fu2020}, 
ultracold atomic optical lattices~\cite{PhysRevLett.93.056402}, photonic-spin waveguides~\cite{PhysRevA.93.062104}, or quantum superconducting circuits ~\cite{Vepslinen2020,Ma2020}. 

%----------------------------------------------------------

\section{Acknowledgements}

The authors thank Tony John George Apollaro and Deniz N. Bozkurt for fruitful discussions.\\

%----------------------------------------------------------
\begin{appendices}

%%%%%%%%%%%%%%%%%%%%%%%%%%%%%%%%%%%%%%%%%%%%%%%%%%%%%%%
\section{Perfect state and entanglement transfer on circulant graphs}\label{sec:PSTonCirculantGraphs}
%%%%%%%%%%%%%%%%%%%%%%%%%%%%%%%%%%%%%%%%%%%%%%%%%%%%%%%

We present
perfect state and entanglement transfer on some circulant graphs in this appendix. Such graphs occupy a relatively large space than linear triangular chains to transfer a state over the same distance and require more qubits to implement.  For
example, we take a \(5 \times 5\) circulant graph shown in ~Fig.\ref{fig:5cycleA} with only nearest-neighbor interactions. The adjacency matrix for such a graph is	
\begin{equation}
	A=\begin{bmatrix}
	0&-\text{i}&0&0&-\text{i}\\
	\text{i}&0&-\text{i}&0&0\\
	0&\text{i}&0&-\text{i}&0\\
	0&0&\text{i}&0&-\text{i}\\
	\text{i}&0&0&\text{i}&0\\
	\end{bmatrix}.
\end{equation} 
\noindent
Fig.~\ref{appFig1} shows that the entanglement transfer on such a graph is nearly perfect with a concurrence of 
\(C\sim 0.93\) at \(t\sim4.5\).

Another example is a pentagram graph which is sketched on ~Fig.\ref{fig:5starB}. This graph contains three triangular plaquettes, but being circulant comes with the cost of more edges. The adjacency matrix reads
\begin{equation}
	A = \begin{bmatrix}
	0&-\text{i}&-\text{i}&-\text{i}&-\text{i}\\
	\text{i}&0&-\text{i}&-\text{i}&-\text{i}\\
	\text{i}&\text{i}&0&-\text{i}&-\text{i}\\
	\text{i}&\text{i}&\text{i}&0&-\text{i}\\
	\text{i}&\text{i}&\text{i}&\text{i}&0\\
	\end{bmatrix}.
\end{equation}
\noindent
Fig.~\ref{appFig2} presents the possibility of nearly perfect entanglement transfer with a concurrence of \(C_{4,5}\sim1\) at \(t\sim3.7\).

%%%%%%%%%%%%%%%%%%%%%%%%%%%%%%%%%%%%%%%%%%%%%%%%%%%%%%%
\section{Eigenvalues and Eigenstates of the Hamiltonian}\label{sec:eigenstates}
%%%%%%%%%%%%%%%%%%%%%%%%%%%%%%%%%%%%%%%%%%%%%%%%%%%%%%%
The eigenstates corresponding to the eigenvalues in Eq.~(\ref{eq:specT}), are listed in the same order as the equations,
\begin{equation}
\begin{split}
\Lambda_1=
\begin{bmatrix}
-0.65 - 0.76 \text{i}\\
 -0.65 + 0.76 \text{i}\\
 0.22 + 0.98 \text{i}\\
 0.22 - 0.98 \text{i}\\
 0 \\
\end{bmatrix},
\Lambda_2=
\begin{bmatrix}
0.65 + 1.15 \text{i}\\
0.65 - 1.15 \text{i}\\
-0.22 + 0.50 \text{i}\\,
-0.22 - 0.50 \text{i}\\
1\\
\end{bmatrix},&\\
\Lambda_3=\begin{bmatrix}
1\\
1\\
1\\
1\\
0 \\
\end{bmatrix},
\Lambda_4=\begin{bmatrix}
 -0.65 + 1.40 \text{i}\\
 -0.65 - 1.40 \text{i}\\
 0.22 + 0.18 \text{i}\\
 0.22 - 0.18 \text{i}\\
 -1\text{i}\\
\end{bmatrix},
\Lambda_5=\begin{bmatrix}
-1.30 + 0.25 \text{i}\\
-1.30 - 0.25 \text{i}\\
0.44 - 0.33 \text{i}\\
0.445 + 0.33 \text{i}\\
1\\
\end{bmatrix}.
\end{split}
\end{equation}

%%%%%%%%%%%%%%%%%%%%%%%%%%%%%%%%%%%%%%%%%%%%%%%%%%%%%%%
\section{Definitions of  the some mathematical and graph theory terms used in the manuscript}\label{sec:footNotes}
%%%%%%%%%%%%%%%%%%%%%%%%%%%%%%%%%%%%%%%%%%%%%%%%%%%%%%%

A graph is a set of vertices and edges connecting them. Here, we present definitions of some mathematical 
terms from graph theory and linear algebra
we used in the main text.

\noindent \emph{Hadamard Product}:Hadamard Product is the element-wise product of two matrices with the same dimensions.\\

\emph{Circulant Graph}:Undirected graphs contain only bidirectional edges. Circulant graphs are undirected graphs, which take any vertex to all of the other vertices~.\\

\emph{Flat Eigenbasis}:When each eigenvector of a basis has entries of the same magnitude, that eigenbasis is called a flat eigenbasis~ \cite{Cameron2014}. \\

\emph{Adjacency Matrix, and the Graph Laplacian Matrix}:A Laplacian matrix $L=D-A$ is a matrix that describes a graph, where $A$ is the adjacency matrix and $D$ is the degree matrix. A degree matrix is a diagonal matrix whose elements indicate the number of edges attached to each vertex of a graph. An adjacency matrix represents
the connections of a graph, whose elements corresponding
to adjacent (connected by an edge) vertices are $1$~\cite{footDegreeM}.

\end{appendices}

\clearpage

%---------------------------------------------------------

\section{References}

\bibliographystyle{unsrt}
\bibliography{ref}

\end{document}